\documentclass[a4paper,11pt]{article}

\pdfoutput=1
\usepackage{comment}
\usepackage{fixltx2e}
\usepackage[table,xcdraw]{xcolor}
\usepackage{jcappub} 
\usepackage{float} 
\usepackage{bm}
\usepackage{stmaryrd}
\usepackage[normalem]{ulem}
\usepackage{amsmath}

\def\be{\begin{equation}}
\def\ee{\end{equation}}
\def\bea{\begin{eqnarray}}
\def\eea{\end{eqnarray}}
\newcommand{\vs}{\nonumber\\}
\def\ba#1\ea{\begin{align}#1\end{align}}
\def\bg#1\eg{\begin{gather}#1\end{gather}}

\newcommand{\s}{\sigma}

\newcommand{\refeq}[1]{Eq.~(\ref{eq:#1})}          
\newcommand{\refeqs}[2]{Eqs.~(\ref{eq:#1})--(\ref{eq:#2})}          
          
\newcommand{\reffig}[1]{Fig.~\ref{fig:#1}}

\newcommand{\refsec}[1]{Sec.~\ref{sec:#1}}          
\newcommand{\refapp}[1]{App.~\ref{app:#1}}
          
\newcommand{\shellPT}[1]{{(#1)_{\rm shell}}}
\newcommand{\renormOP}[1]{\llbracket #1 \rrbracket}

\def\Plin{P_{\rm L}}

\renewcommand{\v}[1]{\bm{#1}}

\renewcommand{\emph}[1]{\textit{#1}}

\newcommand{\vx}{\v{x}}

\newcommand{\vk}{\v{k}}
\newcommand{\vq}{\v{q}}

\newcommand{\vp}{\v{p}}
\newcommand{\vb}{\v{b}}

\newcommand{\<}{\langle}
\renewcommand{\>}{\rangle}
\def\boldlangle{\pmb{\big\langle}}
\def\boldrangle{\pmb{\big\rangle}}

\renewcommand{\d}{\delta}

\newcommand{\Z}{\mathcal{Z}}
\newcommand{\G}{\mathcal{G}}

\newcommand{\Shell}{\mathcal{S}}

\newcommand{\eps}{\epsilon}

\def\L{\Lambda}

\def\leps{\lambda}

\def\SE{\Theta}
\def\se{c}

\def\dlin{\delta^{(1)}}
\def\dlinshell{\dlin_{\rm shell}}

\def\L{\Lambda}

\def\P{\mathcal{P}}
\def\O{\mathcal{O}}
\def\Del{\mathcal{D}}

\newcommand{\perm}[1]{ \expandafter\ifstrempty\expandafter{#1} {\mbox{perm.}} {\mbox{$#1$ perm.}} }

\makeatletter
\newlength{\apb@width}
\newcommand{\autoparbox}[2][c]{\settowidth{\apb@width}{#2}\parbox[#1]{\apb@width}{#2}}
\newcommand{\includegraphicsbox}[2][]{\autoparbox{\includegraphics[#1]{#2}}}
\makeatother

\def\lapl{\nabla^2}
\def\dirac{\delta_{\rm D}}

\renewcommand{\comment}[1]{}

\usepackage[T1]{fontenc}
\usepackage{graphbox}
\usepackage{subfig}
\usepackage{booktabs}
\subheader{TUM-HEP- 1464/23}
\title{Galaxy bias renormalization group}
\usepackage{amsmath}
\author[a]{Henrique Rubira,}
\author[b]{Fabian Schmidt}

\affiliation[a]{Physik Department T31, Technische Universit\"at M\"unchen,\\
James-Franck-Stra{\ss}e 1, D-85748 Garching, Germany}
\affiliation[b]{Max-Planck-Institut f\"{u}r Astrophysik,\\ 
Karl-Schwarzschild-Str. 1, 85748 Garching, Germany}

\emailAdd{henrique.rubira@tum.de}
\emailAdd{fabians@mpa-garching.mpg.de}

\abstract{The effective field theory of large-scale structure allows for a consistent perturbative bias expansion of the rest-frame galaxy density field. In this work, we present a systematic approach to renormalize galaxy bias and stochastic parameters using a finite cutoff scale $\Lambda$. We derive the differential equations of the Wilson-Polchinski renormalization group that describe the evolution of the finite-scale bias parameters with $\Lambda$, analogous to the $\beta$-function running in QFT. We further provide the connection between the finite-cutoff scheme and the renormalization procedure for $n$-point functions that has been used as standard in the literature so far; some inconsistencies in the treatment of renormalized bias in current EFT analyses are pointed out as well. The fixed-cutoff scheme allows us to predict, in a principled way, the finite part of loop contributions which is due to perturbative modes and which, in the standard renormalization approach, is absorbed into counterterms. We expect that this will allow for the robust extraction of (a yet-to-be-determined amount of) additional cosmological information from galaxy clustering, both when using field-level techniques and $n$-point functions.}


\begin{document}

\maketitle
\flushbottom

\section{Introduction}\label{sec:intro}

The large-scale structure of the universe (LSS), as probed via the three-dimensional distribution of galaxies, is known to be a rich source of information on gravity, dark matter, dark energy, and the initial conditions of structure formation. To unlock this information, however, one has to marginalize over the significant uncertainties in our understanding of galaxy formation. What is required in particular is a reliable prediction for the conditional probability of forming a galaxy that passes observational luminosity and color selections, at a given location. The effective field theory (EFT) approach \cite{Baumann:2010tm,carrasco/etal:2012,Carroll:2013oxa,Konstandin:2019bay}, built upon cosmological perturbation theory, allows for a consistent expansion of the galaxy density field into operators $O$, ranked in terms of perturbative order, and corresponding bias coefficients $b_O$ as well as stochastic contributions described by a field $\eps$ and coefficients $c_{\eps,O}$ (see \cite{Desjacques:2016bnm} for a review):
\be
\d_g(\vx,\tau) \equiv \frac{n_g(\vx,\tau)}{\bar n_g(\tau)} - 1
= \sum_O \left[ b_O(\tau) + c_{\eps,O}(\tau) \eps(\vx,\tau) \right] O(\vx,\tau)
+ \eps(\vx,\tau) .
\label{eq:bias}
\ee
The bias parameters and stochastic coefficients are only functions of time $\tau$, while the stochastic field does not correlate with the operators $O$  and is characterized completely by its moments \cite{MSZ}. The bias operators are constructed out of the matter density field $\d(\vx,\tau) = \rho_{\rm m}(\vx,\tau)/\bar\rho_{\rm m}(\tau)-1$. It is a nontrivial result that a local-in-time relation such as \refeq{bias} can also capture the long time scales over which galaxies form \cite{MSZ,senatore:2014} (see also \cite{matsubara:2008}), i.e. despite the fact the theory is fundamentally nonlocal in time. This is a consequence of the fact that, at any given order in perturbation theory, the solutions to the equation of motion for matter and gravitational potential are a sum over a finite number of terms which each are factorizable functions of time and space \cite{matsubara:2015,Schmidt:2020ovm}.

When computing observable statistics such as $n$-point correlation functions based on \refeq{bias} (technically these involve expectation values of products of composite operators), loop integrals arise which have to be regularized. The standard approach has so far been to work at the level of $n$-point correlation functions, and to remove the UV-sensitive loop integrals via counterterms \cite{Assassi2014}; for recent work on different regularization schemes for the galaxy bias, see \cite{Patrone:2023cqe}. In this paper, our aim is instead to follow the renormalization-group (RG) approach of Wilson \cite{wilson:1971} and Polchinski \cite{polchinski:1984}, which proceeds by integrating out small-scale modes in the partition function.\footnote{Two main different RG approaches are used in the context of field theory: the Wilsonian RG and the continuum RG \cite{Schwartz:2014sze}. The Wilsonian RG makes use of the fact that the physics at energy scales $E\ll \L$ has to be independent of the finite cutoff $\L$ (although the couplings may change with $\L$). Later, Polchinski derived the Wilsonian equations in terms of differential increments in the path integral formalism \cite{polchinski:1984}, an approach similar to the one taken in this work. The continuum case instead makes direct use of the invariance of an observable with respect to any arbitrary scale, e.g. via dimensional regularization \cite{StueckelbergdeBreidenbach:1952pwl,Gell-Mann:1954yli,Callan:1970yg, Symanzik:1970rt}. It is more frequently used in the field theory context since it usually allows for more straightforward calculations than the Wilson/Polchinski approach. We will see here however that the latter can be readily applied in the EFT of LSS context.} Ref. \cite{Carroll:2013oxa} was the first to point out how this approach can be adapted to the EFT of LSS. Ref.~\cite{Cabass:2019lqx} used it to derive the EFT prediction for the conditional probability density of the galaxy field given the matter density field; that is, the crucial ingredient needed for making predictions for galaxy clustering, as noted above.

The partition function is usually formulated in terms of a Lagrangian for the fields under consideration. However, the nonlinear dynamics of collisionless matter under gravity cannot be derived from an interaction Lagrangian written in terms of three-dimensional fields. One can reverse-engineer a Lagrangian that gives rise to the fluid equations \cite{matarrese/pietroni,floerchinger/etal}. This Lagrangian will however not give rise to the corrections to a perfect fluid, such as effective pressure and stochastic fluctuations, which are expected in the EFT. It is even less applicable to generate the EFT for biased tracers who do not obey mass and momentum conservation. Instead, our partition function will be based directly on the bias expansion \refeq{bias}, and the free propagator for the linear density field, that is the linear power spectrum $\Plin(k)$.

The Wilson-Polchinski approach to renormalization proceeds by integrating out modes \emph{in the free field} above a momentum cutoff $\Lambda$. In the LSS case, this corresponds to integrating modes above the cutoff \emph{in the linear density field}. 
This is in fact essential in order to keep predictions for the galaxy density field under perturbative control, as noted in \cite{Schmidt:2020viy}.
In this paper, we continue to work along these lines by deriving how the bias coefficients $b_O^\L$ (and stochastic amplitudes), which now depend explicitly on the finite scale $\L$, evolve under a change in the cutoff. Following Wilson's foundational work on this topic, this is known as ``renormalization group flow'' in the field theory context.  

Our results can be useful in several ways:
\begin{itemize}
  \item Observational measurements of bias coefficients $b_O^\L$ at different $\L$ can be compared with the EFT prediction for the running, providing an additional cross-check of EFT-based cosmology inferences based on the same data.
  \item Simulation measurements of bias coefficients are typically performed using $n$-point correlation functions on large scales (e.g., \cite{saito/etal:14}) or the separate-universe technique \cite{lazeyras/etal,baldauf/etal:2015,li/hu/takada:2016}, both of which return bias coefficients in the low-scale limit $\L\to 0$. If one wants to use these measurements to put prior information on finite-scale bias parameters, one needs to run the former to the desired finite $\L$ (see \refsec{running} and \refsec{connection}). 
  \item We point out that the finite-scale renormalization scheme allows us to predict finite contributions to loop integrals that are completely absorbed by counterterms in the approach used so far (\refsec{diffcontrib}). This indicates that there might be more cosmological information to be gained when adopting the finite-scale renormalization instead.
\end{itemize}

The outline of the paper is as follows. We first introduce our notation and conventions. In \refsec{running} we derive the RG equations for the galaxy bias and stochastic parameters, discussing particular solutions for the bias evolution in $\L$. We connect the finite-$\L$ renormalization scheme to the $n$-point renormalized bias in \refsec{connection} and discuss in \refsec{diffcontrib} how the two schemes treat loop contributions. We conclude in \refsec{disc}. The appendices contain details on the more technical calculations. We use a Planck 2018 Euclidean $\Lambda$CDM cosmology  \cite{Planck:2018vyg} for all numerical results.

\subsection*{Dictionary and Notation}

We provide here a brief explanation of the terms and notations used in this paper.

\paragraph{Fourier conventions.}
We keep the letters $\vk,\vp$ and $\vq$ for momenta variables. We use bold letters for three-vectors and subscripts to refer to sum of vectors, e.g. $\vp_{1 \dots n} = \vp_{1} + \dots + \vp_{n}$.  We use the following notation for the Fourier-space integrals
\be
\int_{\vp_1,\dots,\vp_n} = \int \frac{d^3p_1}{(2\pi)^3} \dots \int \frac{d^3p_n}{(2\pi)^3} \,.
\ee
The smoothing of a field $f$ on a length scale $1/\L$ with a filter $W$ is done in real space via
\be
f_{\L}(\vx) = \int d^3x' \,W_{\L}(\vx'-\vx) f(\vx')\,,
\ee
such that in Fourier space
\be
f_{\L}(\vk) = \,W_\L(\vk) f(\vk)\,.
\ee
Throughout this work we assume the filter $W_\L$ to be a sharp filter in Fourier space. 

\paragraph{Scales.}
\begin{itemize}
\item $\L$: denotes the smoothing scale when not fixed to any value; 
\item $\L_\ast$: is the {\it fixed} reference value of the smoothing scale, i.e. the fixed cutoff used in a given analysis; 
\item $k_{\rm NL}$: is the range of validity of the theory, where perturbation theory breaks down; 
\item $k_{\rm max}$: is the highest wavenumber considered in a given analysis.
\end{itemize}
Within the perturbative RG flow context, the hierarchy $k_{\rm max}<\L, \L_\ast<k_{\rm NL}$ has to be fulfilled. 

\paragraph{Operators.}
\begin{itemize}

\item $O[\d]$: General operator, which is constructed as a convolution of density fields 
    \be
    O[\d](\vk) = \int_{\vp_1,\ldots,\vp_n} \dirac(\vk-\vp_{1\ldots n}) S_O(\vp_1,\ldots \vp_n) \d(\vp_1) \cdots \d(\vp_n).
    \label{eq:Odef}
    \ee
    We refer to the number of density fields appearing in the definition as ``order'' of the operator. Hence in case of \refeq{Odef}, the order is $n$, and we also denote this as $O^{[n]}$.

\item We work with the following basis of operators in \refeq{bias}:
  
    \ba
    \mbox{First order:}&\quad \d \,;\vs
    \mbox{Second order:}&\quad \d^2,\  \G_2\,; \vs
    \mbox{Third order:}&\quad \d^3,\  \d\,\G_2,\  \Gamma_3,\  \G_3 \,;
    \ea
    where the Galileon operators are defined in terms of the scaled gravitational potential $\Phi_g \equiv \nabla^{-2}\d$ via
    \bea
    \G_2( \Phi_g) &\equiv& (\nabla_i\nabla_j\Phi_g)^2 - (\nabla^2  \Phi_g)^2 \label{eq:galileon2}\,, \\
    \G_3( \Phi_g) &\equiv& -\frac{1}{2}\left[2\nabla_i\nabla_j  \Phi_g \nabla^j\nabla_k  \Phi_g\nabla^k\nabla^i  \Phi_g + (\nabla^2  \Phi_g)^3-3(\nabla_{i}\nabla_{j} \Phi_g)^2\nabla^2  \Phi_g\right]\,. \label{eq:galileon3}
    \eea
    The operator
    \bea
    \Gamma_3( \Phi_g,\Phi_v) \equiv \G_2(\Phi_g) - \G_2(\Phi_v)
    \eea
    is constructed out of the gravitational, $\Phi_g$, and velocity potential, $\Phi_v \equiv \nabla^{-2} \bm{\nabla}\cdot\v{v}$.
    
    We use the notation $\s^2_{\vk_1,\vk_2}  = \left( \vk_1 \cdot \vk_2/k_1k_2 \right)^2 - 1$ to define $\G_2$ in Fourier space, i.e. [cf. \refeq{Odef} above]
      \be
      S_{\G_2}(\vk_1,\vk_2) = \s^2_{\vk_1,\vk_2}.
    \ee
    
  \item The evolved matter density field $\d$ is expanded in terms of the linear field $\dlin$ via
    \be
    \d(\vk) = \sum_{\ell=1}^\infty\int_{\vp_1,\ldots,\vp_\ell} \dirac(\vk-\vp_{1\ldots \ell}) F_\ell(\vp_1,\ldots \vp_\ell) \dlin(\vp_1) \cdots \dlin(\vp_\ell),
    \label{eq:dexp}
    \ee
    with $F_\ell$ being the usual PT kernels \cite{lssreview}, including $F_1 \equiv 1$.
    
  \item We denote the contribution to a given operator $O$ at $m$-th order as $O^{(m)}$, so if $O = O^{[n]}$, then $m\geq n$. In other words, superscripts in parenthesis refer to the \emph{expansion order} of an operator and superscripts in square brackets refer to the \emph{order} of an operator.
    
    \item   $O[\d_\L^{(1)}]$: same operator as \refeq{Odef} above, but constructed by smoothing the initial conditions at a scale $\L$. Note that we use the shorthand notation $O[\d_\L^{(1)}] \equiv O[\d[\d_\L^{(1)}]]$ for operators constructed from the forward-evolved matter density field, which in turn is initialized with the cut linear density field. This should not be understood as operators directly constructed from the linear density field. 
    
  \item $\renormOP{O}$: the $n$-point  function renormalized operator $O$ introduced by \cite{Assassi2014} and defined using the renormalization conditions, for all $m\geq0$, 
    \be
    \langle \d^{(1)}(\vk_1)\cdots\d^{(1)}(\vk_m) \renormOP{O}(\vk) \rangle
    \stackrel{k_i \to 0}{\longrightarrow}
    \langle \d^{(1)}(\vk_1)\cdots\d^{(1)}(\vk_m) O[\d](\vk) \rangle_{\rm LO}\,,
    \label{eq:renormcond_pre}
    \ee
    where LO indicates the leading order term (see \refsec{connection}). The renormalized $\d^2$ operator, $\renormOP{\d^2}$,  at 1-loop and up to $m=2$ reads
    \be
    \renormOP{\d^2} = \d^2 -\s^2_\infty \left( 1 + \frac{68}{21}\d + \frac{8126}{2205}\d^2 + \frac{254}{2205}\G_2 \right) \,,
    \label{eq:renorm_d2}
    \ee
    where $\s^2_\infty = \lim_{\L\to\infty}\s^2_\L$ [see \refeq{sigmaLambda}].
    For a broader discussion on different renormalization schemes, see \cite{Patrone:2023cqe}.

  \item We expand the power spectrum of the stochastic field $\eps$ in \refeq{bias} as (cf. \cite{Desjacques:2016bnm})
    \be
    \< \eps(\vk) \eps(\vk')\> = (2\pi)^3 \dirac(\vk+\vk') \sum_{n=0}^\infty P_\eps^{\{2n\}} k^{2n},
  \label{eq:Pstochdef}
  \ee
  where we also often abbreviate $P_\eps \equiv P_\eps^{\{0\}}$. Note that $\< \eps(\vk) \eps(\vk')\>$ corresponds also to the stochastic contribution to the galaxy power spectrum.  
\end{itemize}

\paragraph{Partition function.}
\begin{itemize}
    \item $\P[\varphi]$: The probability distribution function (PDF) of a field $\varphi$, which enters as a measure of the path integral. If $\varphi$ is assumed to be Gaussian, then 
    \be
        \P[\varphi] = \left(\prod_{\vk}^\L 2\pi P_{\varphi}(k)\right)^{-1/2} \exp\left[-\frac12 \int_{\vk}^\L \frac{|\varphi|^2}{P_{\varphi}(k)}\right]\,, \label{eq:PlikeDef}
    \ee
    where $P_{\varphi}$ is the power spectrum of $\varphi$.
    \item $\Z [J_\L]$ : The partition function evaluated on a current with support over wavenumbers up to $\L$. We have that  
    \be \label{eq:Z_intro}
        \Z [J_\L] = \int \Del\dlin_\L \, \P[\dlin_\L] \exp\left(S_{\rm int}[\dlin_\L,J_\L]  \right)\,,
    \ee
    where $S_{\rm int}[\dlin_\L,J_\L]$ is the interaction action. 
\end{itemize}

\paragraph{Initial conditions.} We assume Gaussian initial conditions throughout [cf.~\refeq{PlikeDef}]. We keep only the fastest growing mode, allowing us to phrase the initial conditions completely in terms of $\varphi=\dlin$. 
We consider three types of power spectra (or propagators) for $\dlin$: the linear propagator
    \be
    \Plin(k) = \< \d^{(1)}(\vk)\d^{(1)}(\vk')\>^{'} ,
    \ee
    the linear propagator cut at $\L$ 
    \be
    \Plin^\L(k) = \< \d^{(1)}_\L(\vk)\d^{(1)}_\L(\vk')\>^{'}  = \< \d^{(1)}_\L(\vk)\d^{(1)}(\vk')\>^{'} ,
    \ee
    where we used the sharp-$k$ cutoff properties in the second equality, and the shell propagator 
    \be \label{eq:Pshell}
    P_{\rm shell}(k) = \< \dlinshell(\vk)\dlinshell(\vk')\>^{'}
    \ee
    which only has support within an infinitesimal shell $[\L,\L+\leps]$ in momentum space. Throughout, we use the primed notation,
      \be
      \<O(\vk_1) \dots O(\vk_n)\> = (2\pi)^3\, \dirac(\vk_{1\dots n})\,\<O(\vk_1) \dots O(\vk_n)\>'
      \ee
      for ensemble averages with the overall momentum constraint removed.
    
    We also define the variance of the linear density field including modes up to $\L$ as $\s^2(\L)$, i.e. 
    \be
    \sigma^2_\L  = \int_{\vp}^\L \Plin(p) .
    \label{eq:sigmaLambda}
    \ee  

\paragraph{Feynman rules.}
\begin{itemize}
\item The different propagators are denoted via
    \ba
    \raisebox{-0.0cm}{\includegraphicsbox[scale=1]{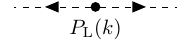}} 
    \qquad \raisebox{-0.0cm}{\includegraphicsbox[scale=1]{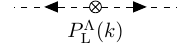}} \qquad
    \raisebox{-0.0cm}{\includegraphicsbox[scale=1]{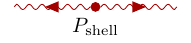}} . \nonumber
    \ea

  \item There are two types of vertices: large boxes representing the operator convolution \refeq{Odef}  and small boxes indicating the expansion of $\d$ in terms of linear legs \refeq{dexp}, i.e.,
    \be
    \raisebox{-0.0cm}{\includegraphicsbox[scale=1]{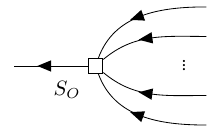}}
    \raisebox{-0.0cm}{\includegraphicsbox[scale=1]{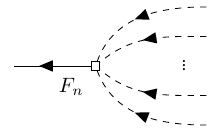}}
    \quad.
    \ee
    Notice that the linear legs are always represented by dashed lines, to make the distinction between both vertices clearer.
\end{itemize}

\section{RG flow via the partition function} \label{sec:running}

In this section we compute the RG running of the bias parameters using the partition function formalism.
The partition function for matter can be written as \cite{Carroll:2013oxa,Cabass:2019lqx} 
\ba
\Z^{\rm m}[J_{\L}] &= \int \Del\dlin_\L \P[\dlin_\L] \exp\left(\int_{\vk} J_{\L}(\vk) \d_{\rm fwd}[\dlin_\L](-\vk)\right)\,, \label{eq:ZofJm} \,
\ea
where $\P[\dlin_\L]$ is defined in Eq.~(\ref{eq:PlikeDef}) and $\d_{\rm fwd}[\dlin_\L]$ denotes the perturbative (forward-evolved) solution to the equations of motion for the matter density field.
As discussed in \refsec{intro}, we have to insert the solution of the equations of motion in the coupling to the source current $J_{\L}$, which, as indicated by the subscript, only has support over wavenumbers below $\L$. Moreover, we only keep the fastest-growing mode, allowing us to reduce the initial conditions to a single field $\dlin$. Notice that we already normalized the partition function in $\P[\dlin_\L]$ [see \refeq{PlikeDef}], i.e. $\Z^{\rm m}[0] = 1$.

The partition function generates all $n$-point correlation functions of the matter density field via functional derivatives with respect to $J_{\L}$ \cite{Carroll:2013oxa}. 
The finite cutoff ensures that all observables derived from $\Z$ are finite, but of course accurate only below $\L$, which is why we consider a source current cut at $\L$. However, since the cutoff is imposed artificially as a calculational tool, observables should not depend on $\L$. In order to satisfy this
requirement in general, one has to introduce counterterms whose $\L$-dependent coefficients absorb the cutoff dependence. Physically, the counterterms capture the effect of small-scale modes that have been integrated out to obtain \refeq{ZofJm}. In the case of the matter density, the leading such counterterm is $c_{\nabla^2\d}(\L) \nabla^2\d$ \cite{Baumann:2010tm}.

We now turn to the corresponding partition function for biased tracers. It can be similarly guessed from the form of the bias expansion \refeq{bias} to be \cite{Cabass:2019lqx}
\ba
\Z[J_{\L}]
= \int \Del\dlin_\L \P[\dlin_\L] \exp&\left(\int_{\vk} J_{\L}(\vk) \left[\sum_O b_O^\L O[\dlin_\L](-\vk)\right]  \right. \vs
&\quad \left. + \frac12 P_\eps^\L \int_{\vk} J_{\L}(\vk) J_{\L}(-\vk)  + \O[J_\L^2 \dlin_\L,\  J_\L^3]  \right)\,. \label{eq:ZofJ} 
\ea
There are thus two differences to \refeq{ZofJm}. First, we have included the counterterms, including their $\L$-dependent bias coefficients $b_O^\L$, since they already appear at linear order and lowest order in derivatives for biased tracers. 
Second, we have a term $\propto J_{\L}^2$, which captures stochasticity (noise) in the tracer field \cite{Cabass:2019lqx}. Specifically, we include the leading Gaussian part  $P_\eps^\L \equiv P_\eps^{\{0\},\L}$ [cf. \refeq{Pstochdef}]; this is proportional to the variance of the field $\eps(\vx,\tau)$ in \refeq{bias}]. On the other hand, we have not written the stochastic terms that incorporate the coupling with the operators $O[\dlin_\L]$, corresponding to the $c_{\eps,O}$ in \refeq{bias}. We also have not explicitly written \emph{non-Gaussian} stochastic terms which are of order $J_{\L}^3$ and higher \cite{Cabass:2020nwf}, relevant, for instance, for the bispectrum. We will neglect these terms throughout the paper as they are not relevant for the running of the bias parameters.
Their investigation is left for future work.
Note that a noise contribution exists also for the matter density, which we have not written in \refeq{ZofJm}, as it is suppressed by $k^4$, and hence only relevant at fairly high order in perturbation theory.

Similarly to \refeq{ZofJm}, \refeq{ZofJ} generates all $n$-point correlation functions of the galaxy density field $\d_g$. 
In addition, it can be used to derive the probability density functional (or ``likelihood'') of the tracer field \cite{Cabass:2019lqx}.

\subsection{General RG equations} \label{sec:generalRG}  

We start by reviewing the results from \cite{Carroll:2013oxa}, who derived the running of the matter counterterms using the partition function.  They expand the action in \refeq{ZofJm} as a general series in $\dlin$ and $J$, introducing index notation for momentum vectors and Einstein summation convention:
\be
S_{\rm int}[\dlin_\L, J^\L] = \sum_{n=1}^\infty\sum_{m=0}^\infty \frac1{n!m!}K^{j_1\ldots j_n}_{i_1\ldots i_m}(\L) (\dlin_\L)^{i_1}\cdots (\dlin_\L)^{i_m}\; J^\L_{j_1}\ldots J^\L_{j_n}.
\ee
We have already assumed only the fastest growing mode, allowing us to reduce the degrees of freedom from $\phi=(\d,\theta)$ to $\d$.
Here, the terms with $n=1$ correspond to the ``deterministic'' (mean-field) prediction for $\d_g$, i.e. the first line of \refeq{ZofJ}, while contributions with $n>1$, i.e. those nonlinear in the current $J_\L$, encode stochasticity as noted above. Each of the kernels $K$ is a sum of several terms involving contributions from several operators at different orders in perturbations. 
In fact, this expansion of the effective action and the RG equations derived from it are directly applicable to the case of biased tracers [\refeq{ZofJ}] as well. 
Thus, the kernels include the bias and stochastic parameters from the large-scale bias expansion, but defined at the scale $\L$.

Ref.~\cite{Carroll:2013oxa} then derive RG equations for the coefficients
$K^{j_1\ldots j_n}_{i_1\ldots i_m}(\L)$ [Eq.~(4.22) there]. For a kernel of order $n$ (upper indices),
the source terms in the RG equation
only involve orders $\leq n$ (this justifies why we can drop non-Gaussian stochastic terms in this paper). Since kernels with $n=0$ vanish (for Gaussian initial conditions, at least), the RG equations for the $n=1$ (bias) and $n=2$ (Gaussian stochastic) kernels become 
\ba
\frac{d}{d\L} K^j_{i_1\ldots i_m} &= -\frac12 \frac{d (\Plin^\L)^{kl}}{d\L} K^j_{kli_1\ldots i_m} \label{eq:RGgeneral1}\\
\frac{d}{d\L} K^{j_1j_2}_{i_1\ldots i_m} &= -\frac12 \frac{d (\Plin^\L)^{kl}}{d\L} K^{j_1j_2}_{kli_1\ldots i_m} + 2 \frac{d (\Plin^\L)^{kl}}{d\L} \sum_{k=0}^m
{m \choose k}
K^{j_1}_{ki_1\cdots i_k} K^{j_2}_{li_{k+1}i_m} ,
\label{eq:RGgeneral2}
\ea
where $d(\Plin^\L)^{kl}/d\Lambda = \delta_{\rm K}^{kl} P_{\rm shell}$ is the shell propagator written in index notation.
Note that, if and only if the bias expansion is complete at each order in perturbations, the kernels $K$ depend on $\L$ only through the coefficients $b_O^\L$. 
In the next section, we will derive the RG equations explicitly at the level of bias operators, show that indeed the $b_O^\L$ completely absorbs the running of the kernels, and connect the results with the general RG equation above.

\subsection{RG equations for biased tracers}

The partition function $\Z[J]$ defined at some scale $\L'$ predicts the running of the bias coefficients $b_O^{\L'}$ to any lower scale $\L < \L'$. To compute this, we
consider the running over an infinitesimal distance from $\L'$ to $\L = \L' - \leps$. This corresponds to integrating out modes in a \emph{shell} of width $\leps$ in momentum space. In RG flow computations, it is common to do this by computing the derivative of $\Z$ with respect to $\L$; here we begin by considering a finite cutoff shift $\leps$, which is assumed to be infinitesimal afterwards.

The computation of the running proceeds by integrating out the modes of the linear density field within the shell $|\vk| \in (\L,\L']$. Correspondingly, we separate $\dlin_{\L'}$ into two parts:
\be \label{eq:shelldef}
\dlin_{\L'}(\vk) = \dlin_\L(\vk) + \dlinshell(\vk),
\ee
where $\dlinshell(\vk)$ has support $|\vk| \in (\L,\L']$, and its power spectrum $P_{\rm shell}$ is defined in \refeq{Pshell}.
The path integral in \refeq{ZofJ} factorizes, leading to
\ba
&\Z[J_{\L}] = \int \Del\dlin_\L \P[\dlin_\L] \int \Del\dlinshell \P[\dlinshell] \label{eq:Z_shell}  \\
&\quad \times\exp\left(\int_{\vk} J_{\L}(\vk) \left[\sum_O b_O^{\L'} O[\dlin_\L+\dlinshell](-\vk)\right] + \frac12 P_\eps^{\L'} \int_{\vk} J_{\L}(\vk) J_{\L}(-\vk)  + \O[J_\L^2 \dlin_\L,\  J_\L^3] \right) \,. \nonumber
\ea
We emphasize again that $J_{\L}$ has support only at scales below $\L$, while the right-hand side integrates over all modes $\dlin_{\L'} = \dlin_{\L+\leps}$. Thus, $J_\L$ has no support in the shell, which guarantees that the product $P_{\rm shell}(k) J_{\L}(k)$ is zero (in other words, $J_\L$ and $\dlinshell$ are orthogonal). If we are interested in computing observables up to a maximum scale $k_{\rm max}$, we can in fact choose $J$ to have support only up to $k_{\rm max}$, and choosing $\L' > \L > k_{\rm max}$ is sufficient to fulfill this condition (where all of these scales are below $k_{\rm NL}$).
In perturbation theory, every operator $O$ is evaluated up to some finite maximum order. Let us consider the $n$-th order contribution $O^{(n)}$, which involves $n$ powers of the linear density field (including of course the contributions from the nonlinear evolution of matter itself). Thus, we can write
\bea \label{eq:Oshell}
O^{(n)}[\dlin_\L+\dlinshell] &=& O^{(n)}[\dlin_\L] + O^{(n),\shellPT{1}}[\dlin_\L, \dlinshell] + O^{(n),\shellPT{2}}[\dlin_\L, \dlinshell] \\
&& \qquad\qquad\qquad \qquad  \qquad + \ldots + O^{(n),\shellPT{n-1}}[\dlin_\L, \dlinshell] + O^{(n)}[\dlinshell], \nonumber
\eea
where the $O^{(n),\shellPT{m}}$ involve $m$ powers of $\dlinshell$ and $n-m$ powers of $\dlin_\L$. Trivially in this notation $O^{(n),\shellPT{0}}=O^{(n)}[\dlin_\L]$ and $O^{(n),\shellPT{n}}=O^{(n)}[\dlinshell]$. We refer the reader to App.~\ref{app:shellOP} for examples on how this expansion works for the basis of operators considered in this paper.

  Note that in the limit $\leps\to 0$, $\dlinshell$ is parametrically smaller than $\dlin_\L$; to see this, note that $\<\dlinshell(\vx)^2\> = (d\s_\L^2/d\L) \leps$ [see \refeq{sigmashell} below], which in the limit $\leps\to 0$ is much smaller than  $\<\dlin_\L(\vx)^2\> = \s_\L^2$.
  Using this, we can expand the exponential in Eq.~(\ref{eq:Z_shell}) in powers of $\dlinshell$:
\ba \label{eq:Z_S}
  \Z[J_{\L}] &= \int \Del\dlin_\L \P[\dlin_\L] \exp\left(\int_{\vk} J_{\L}(\vk) \left[\sum_O b_O^{\L'} O[\dlin_\L](-\vk)\right]\right. \vs
 &\qquad\qquad\left. + \frac12 P_\eps^{\L'} \int_{\vk} J_{\L}(\vk) J_{\L}(-\vk) + \O[J_\L^2 \dlin_\L,\  J_\L^3] \right) \vs
&\qquad\times 
\bigg( 1 + \int_{\vk} J_{\L}(\vk) \left[\sum_O b_O^{\L'} \left(  \Shell_{O}^{1} [\dlin_\L](-\vk) +  \Shell_{O}^{2} [\dlin_\L](-\vk) + \ldots \right) \right] 
\vs
&\qquad \qquad+ \frac12 \int_{\vk,\vk'}  J_{\L}(\vk) J_{\L}(\vk')  \sum_{O, O'} b_O^{\L'} b_{O'}^{\L'}  \left[\Shell_{OO'}^{11} [\dlin_\L] (\vk,\vk') + \ldots \right] + \O[J_\L^2 \dlin_\L,\  J_\L^3] \bigg),   
\ea
where we introduce the operator expectation values
\ba
\Shell_{O}^{2} [\dlin_\L] &= \sum_{n\geq 2} \int \Del\dlinshell \P[\dlinshell] \: O^{(n),(2)_{\rm shell}} [\dlin_\L,\dlinshell](\vk) 
\vs
&= \sum_{n\geq 2} \left\< O^{(n),(2)_{\rm shell}} [\dlin_\L,\dlinshell](\vk) \right\>_{\rm shell}
\label{eq:shell}\\
\Shell_{OO'}^{11} [\dlin_\L] (\vk,\vk') &= \sum_{n,n'\geq 1} \int \Del\dlinshell \P[\dlinshell] \: O^{(n),(1)_{\rm shell}} [\dlin_\L,\dlinshell](\vk)\: O'^{(n'),(1)_{\rm shell}} [\dlin_\L,\dlinshell](\vk') 
\vs
&= \sum_{n,n'\geq 1} \left\< O^{(n),(1)_{\rm shell}} [\dlin_\L,\dlinshell](\vk)\: O'^{(n'),(1)_{\rm shell}} [\dlin_\L,\dlinshell](\vk') \right\>_{\rm shell} \,,
\nonumber
\ea
where the sums run over the contributions to $O$ and $O'$ in perturbation theory. On the other hand,
\be
\Shell_{O}^{1} [\dlin_\L] = \sum_{n\geq 1} \int \Del\dlinshell \P[\dlinshell] \: O^{(n),(1)_{\rm shell}} [\dlin_\L,\dlinshell](\vk) = 0
\ee
vanishes, since the integrand is linear in $\dlinshell$ by definition.
We also emphasize that these expectation values are functionals of $\dlin_\L$; that is, the long-wavelength modes are \emph{not} integrated out.
Terms that are proportional to more than two powers of $\dlinshell$, represented in \refeq{Z_S} by dots, are suppressed such that we drop them hereafter. Thus, only $S_O^2$ and $S_{OO'}^{11}$ need to be considered.

\begin{figure}[t]
    \centering
    \includegraphics[width = 0.3\textwidth]{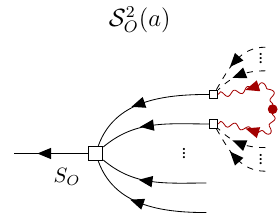}
    \includegraphics[width = 0.3\textwidth]{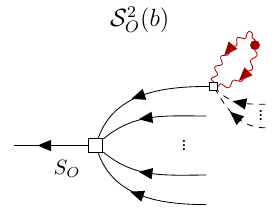}
    \includegraphics[width = 0.6\textwidth]{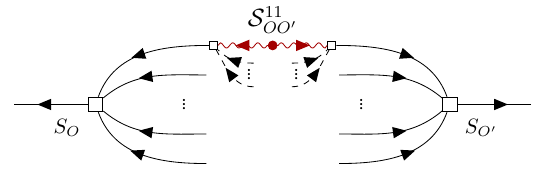}
    \caption{Diagrams representing the shell correlators $\Shell_O^2$ (top) and $\Shell_{OO'}^{11}$ (bottom). }
    \label{fig:Sdiag}
\end{figure}

We can then reconstruct the exponential from its series and compare it to the partition function for the bias and stochastic parameters at the scale $\L$:
\ba
\Z[J_{\L}] &= \int \Del\dlin_\L \P[\dlin_\L] \exp\left[ \int_{\vk} J_{\L}(\vk) \left(\sum_O b_O^{\L'} \left[ O[\dlin_\L](-\vk) + \Shell_{O}^{2} [\dlin_\L](-\vk) \right] \right) \right. \label{eq:ZS1}\\
  & \left.\qquad +  \frac12 P_\eps^{\L'} \int_{\vk,\vk'} J_{\L}(\vk) J_{\L}(\vk') \left( (2\pi)^3\dirac(\vk+\vk') +  \sum_{O, O'} b_O^{\L'} b_{O'}^{\L'}  \Shell_{OO'}^{11} [\dlin_\L](\vk,\vk')  \right) \right] \,. \nonumber
\ea
Fig.~\ref{fig:Sdiag} provides a diagrammatic representation for the shell correlators $\Shell_O^2$ (top) and $\Shell_{OO'}^{11}$ (bottom). 
We can also connect these results to the general RG equations in \refeqs{RGgeneral1}{RGgeneral2}, by noting that $\mathcal{S}_O^{2} (\vk_1,\ldots \vk_m)$ corresponds to the contribution of the operator $O$ to 
\be
\frac{d (\Plin^\L)^{kl}}{d\L} K^j_{kli_1\ldots i_m}\Big|_O\,,
\ee
and correspondingly $\mathcal{S}_{OO'}^{11} (\vk_1,\ldots \vk_m)$ is the contribution of operators $O,O'$ to 
\be
\frac{d (\Plin^\L)^{kl}}{d\L} K^{j_1}_{ki_1\cdots i_k}\Big|_O K^{j_2}_{li_{k+1}i_m}\Big|_{O'}\,.
\ee

 Note that $S_{OO'}^{11}$ multiplies two powers of the current $J_\L$. This means that it contributes to the stochastic part of the partition function, in particular to the coupling between $\O(J_\L^2)$ with long-wavelength operators $O[\dlin_\L]$. Since we disregard these higher-order stochastic contributions in this paper, we do not need to consider $S_{OO'}^{11}$ further. We note though that it is ultimately these stochastic terms which also produce the leading stochastic term $P_\eps^\L \int_{\vk} J_\L(\vk) J_\L(-\vk)$ under the RG flow. 

Let us then focus on the terms linear in $J_\L$. 
By comparing \refeq{ZS1} with the partition function \refeq{ZofJ} at the scale $\L$, we can read off
\be
\sum_O b_O^\L O[\dlin_\L] = \sum_O b_O^{\L'} \left(O[\dlin_\L] + \Shell_O^2[\dlin_\L]\right) .
\label{eq:match}
\ee
If (and only if) the set of bias operators is closed under renormalization, the shell expectation value $\Shell_O^2[\dlin_\L]$ can depend on $\dlin_\L$ only through the bias operators themselves. That is, we should be able to write
\be
\Shell_O^2[\dlin_\L] = \leps \sum_{O'} \SE_{O,O'}(\L) O'[\dlin_\L],
\label{eq:SO2exp}
\ee
where we have used the fact that $\Shell_O^2[\dlin_\L]$ is first order in the shell width $\leps$. We can anticipate that
\be
\Shell_O^2[\dlin_\L] \propto \int_{\vp} P_{\rm shell}(p)\,,
\ee
and we will show this explicitly in the next section. Since [cf. \refeq{sigmaLambda}]
\be
\int_{\vp} P_{\rm shell}(p) = \int_{\L}^{\L+\leps} \frac{p^2 dp}{2 \pi^2 }\Plin(p) = \frac{d \s^2_\L}{d \L}\Big|_{\L}\leps + \O(\leps^2)\,,
\label{eq:sigmashell}
\ee
we can isolate the leading $\L$ dependence in $\SE$ as 
\be
\SE_{O,O'}(\L') = \se_{O,O'}(\L) \frac{d\s^2_\L}{d\L}\Big|_{\L}\,,
\label{eq:sedef}
\ee
where $\se_{O,O'}(\L)$ are coefficients which will be determined below.
  For operators $O$ that are leading-order in derivatives, $\se_{O,O'}$ turn out to be $\L$-independent coefficients of order unity. For higher-derivative operators $O$, they are $\L$-dependent, with the powers of $\L$ determined by the dimensions of the operator.

Inserting \refeq{SO2exp} into \refeq{match}, and reordering the sum, we obtain
\be
\sum_O b_O^\L O[\dlin_\L] = \sum_O \left[ b_O^{\L'} + \leps \sum_{O'} \SE_{O'O}(\L) b_{O'}^{\L'}\right] O[\dlin_\L] \,.
\ee
Now we can identify coefficients and read off
\be
b_O^\L = b_O^{\L'} + \leps \sum_{O'} \SE_{O'O}(\L) b_{O'}^{\L'}\, ,
\ee
which via \refeq{sedef} and since $\L = \L'-\leps$ becomes 
\be \label{eq:b_diffnew}
\frac{d}{d\L} b_O^\L =
-\frac{d}{d\leps} b_O^\L =
- \sum_{O'} \SE_{O'O}(\L) b_{O'}^{\L} = - \frac{d\s^2_\L}{d\L} \sum_{O'} \se_{O'O}\, b_{O'}^{\L}\, .
\ee
In order to derive the RG equations for the $b_O^\L$, we thus only need to determine the coefficients $\se_{O'O}$. 

\subsection{Running of the bias parameters} \label{sec:bias_running}

We now compute the running of the bias parameters, evaluating the ensemble average of correlators over the initial modes in the shell described by Eq.~(\ref{eq:shell}), \emph{keeping the modes $|\vk| < \L$ fixed}. 
	This is a central difference to the $n$-point-function-based renormalization from \cite{Assassi2014}: our approach keeps \emph{all linear modes below the cutoff explicit}, whereas the $n$-point-function-based renormalization marginalizes over all modes, including large scales.

        We evaluate the running of linear and second-order bias parameters considering up to third-order corrections to those operators.
As argued above, the only non-trivial contributions are the $\Shell^2_O$ terms
\bea 
 \Shell_{O}^{2} [\dlin_\L] (\vk)  &=&  \< O^{\shellPT{2}}  [\dlin_\L,\dlinshell](\vk_1)  \>_{\rm shell}\,,\label{eq:shell2}\,
\eea
with contributions $\Shell_{O}^{\geq 3}$ being suppressed by $\dlinshell$. We now proceed to calculate those shell ensemble averages for operators that are up to the third order in the bias expansion, with the goal of computing the coefficients in the expansion \refeq{SO2exp}. We refer to App.~\ref{app:shellInt} for the complete calculation, and highlight in the main text the most relevant results. 
We also comment later on how fourth-order operators affect the RG running of second-order bias operators. 

The shell integrals for $\Shell^2_\d$ leads to terms that are proportional to $ \int_{\vp} F_\ell(\vp,-\vp,\dots)  P_{\rm shell}(p)$; specifically,
\bea
\Shell^2_{\d} [\dlin_\L](\vk) 
&=&  3\dlin_\L(\vk) \int_{\vp} F_3(\vp,-\vp,\vk)  P_{\rm shell}(p) 
\vs
&& \quad + \,6 \int_{\vp,\vp_1,\vp_2} \dirac(\vk-\vp_{12}) F_4(\vp,-\vp,\vp_1,\vp_{2})  P_{\rm shell}(p) \dlin_\L(\vp_1)\dlin_\L(\vp_{2}) \vs
&& \quad +\, \O\left[\left(\dlin_\L\right)^3\right] \vs 
&\stackrel{k\ll p}{=}&  -\frac{61}{315} k^2 \dlin_\L(\vk) \int \frac{p^2 dp}{2\pi^2} \,   \frac{P_{\rm shell}(p)}{p^2} +\, \O\left[\nabla^2 \left(\dlin_\L\right)^2\right]  \,,
\eea
where $\O[\nabla^2(\dlin_\L)^2]$ represents higher-derivative contributions that are also higher order in $\dlin_\L$. After integrating out modes in the shell, the linear operator $\d$ thus only contributes a correction to the $\nabla^2 \d$ operator. Diagrammatically, higher-derivative operators appear from shell contractions on the same $F_\ell$ kernel, represented by the right top panel of Fig.~\ref{fig:Sdiag}. This is due to the ``double-softness'' of the $F_\ell$ kernels, inherited from mass and momentum conservation \cite{abolhasani/etal}. 
In the case of $\d$, this is the only type of contraction that exists, leading to the conclusion that $\d$ can only source higher-derivative terms.

The shell contraction for the operator $\d^2$ leads to
\ba \label{eq:shell_d2}
  \Shell^2_{\d^2} [\dlin_\L](\vk) &=   4\,\dlin_\L(\vk) \int_{\vp} F_2(\vk,\vp)  P_{\rm shell}(p) \\
  &+  6 \int_{\vp,\vp_1,\vp_2} \dirac(\vk-\vp_{12}) \left[ F_3(\vp,-\vp, \vp_1) +  F_3(\vp,\vp_1, \vp_2)\right] P_{\rm shell}(p) \dlin_\L(\vp_1)\dlin_\L(\vp_{2}) \vs
  &+ 4 \int_{\vp,\vp_1,\vp_2} \dirac(\vk-\vp_{12}) F_2(\vp,\vp_1)F_2(-\vp,\vp_2)  P_{\rm shell}(p) \dlin_\L(\vp_1)\dlin_\L(\vp_{2}) \vs
  &+ \O\left[\left(\dlin_\L\right)^3\right] \,. \nonumber
  \ea
  Here and throughout we assumed $\vk\neq0$, so that we can drop tadpole contributions such as $(2\pi)^3\dirac(\vk) \int_{\vp} P_{\rm shell}(p)$. 
The first term leads to a contribution to $\d$ and the term proportional to $F_3(\vp,-\vp, \vp_1)$ leads to corrections to $\nabla^2 \d$. 
The terms proportional to $F_2(\vp,\vp_1)F_2(-\vp,\vp_2)$ and $F_3(\vp,\vp_1, \vp_2)$ lead to corrections of the type $(\vp_1\cdot \vp_2)/(p_1^2 p_2^2)$, which correspond to non-Galilean-invariant displacements $\partial_i \Phi_g \partial_j \Phi_g$. When summed together, however, those contributions cancel out, similarly to what was found by \cite{Assassi2014}. The sum of all contributions in Eq.~(\ref{eq:shell_d2}) leads to
\ba \label{eq:Sd2}
 \Shell^2_{\d^2}[\dlin_\L] (\vk) = &\left[ \frac{68}{21}\d^{(1+2)}(\vk) + \frac{8126}{2205}\left[\d^{2}(\vk)\right]^{(2)} +\frac{254}{2205}\G_2^{(2)}(\vk) \right] \int \frac{p^2 dp}{2 \pi^2 }P_{\rm shell}(p) \vs 
 &\quad + \mbox{higher derivative (h.d.)} + \O\left[\left(\dlin_\L\right)^3\right]\,, 
\ea
where we omitted higher-derivative (h.d.) and higher-order operators. Notice that \refeq{Sd2} has the same structure and coefficients as Eq.~(2.44) of \cite{Assassi2014} [once one converts $\G_2$ to $(\partial_i\partial_j \Phi_g)^2$ via \refeq{galileon2}]\footnote{Although the numerical coefficients are the same, the frameworks are essentially different. Ref.~\cite{Assassi2014} calculates those coefficients setting a renormalization condition in the IR (see \refsec{connection}), which ultimately leads to contractions with linear legs. In our framework, we integrate out a UV shell based on \cite{Carroll:2013oxa}. Differences appear for instance via non-linear terms when including primordial non-Gaussianities which will generate non-linear terms in the running.}; the only difference is that we integrate over an infinitesimal shell at finite cutoff $\L$, while \cite{Assassi2014} integrate over all modes up to $\L$ and send $\L\to\infty$. Non-trivially, notice that the prefactor in front of $\d^{(1)}(\vk)$ and $\d^{(2)}(\vk)$ are the same, despite the fact that one comes from the first term in Eq.~(\ref{eq:shell_d2}) and the other comes from the sum of $F_2(\vp,\vp_1)F_2(-\vp,\vp_2)$ and $F_3(\vp,\vp_1, \vp_2)$.
This is a direct consequence of the Galilean invariance of the underlying physics: if $\d^{(2)}$ received a different contribution than $\d^{(1)}$, then the running of $\Lambda$ would force us to introduce a new bias term for the displacement term $s^i\partial_i \d$ which is part of $\d^{(2)}$. This term is not Galilei-invariant however and should not appear in the bias expansion.
Another way to say this is to note that the bias coefficients at fixed scale $\Lambda$ are physical parameters, and so their running has to be independent of the order in perturbation theory that we work in. 
In conclusion, we see that $\Shell^2_{\d^2}$ sources $\d,\,\d^2$ and $\G_2$.

The $\d^3$ contribution leads to
\bea \label{eq:Sd3}
\Shell^2_{\d^3}[\dlin_\L] (\vk) &=&  3\dlin_\L(\vk) \int_{\vp} P_{\rm shell}(p)+ 3 \int_{\vp,\vp_1,\vp_2}\dirac(\vk-\vp_{12})  F_2(\vp_1,\vp_2)  P_{\rm shell}(p) \dlin_\L(\vp_1)\dlin_\L( \vp_{2}) \vs
&& + 3 \int_{\vp,\vp_1,\vp_2} \dirac(\vk-\vp_{12}) F_2(\vp_1,\vp)  P_{\rm shell}(p) \dlin_\L(\vp_1)\dlin_\L(\vp_{2}) +  \O\left[\left(\dlin_\L\right)^3\right]\\
&=&  3 \left(\d[\d_\L^{(1)}]\right)^{(1+2)}(\vk)  \int_{\vp} P_{\rm shell}(p) + \frac{17}{7}  \left(\d^2[\d_\L^{(1)}]\right)^{(2)}(\vk)   \int_{\vp}   P_{\rm shell}(p) +  \O\left[\left(\dlin_\L\right)^3\right] \,, \nonumber
\eea
where we see that the first two terms contribute to $\d[\d_\L^{(1)}]$ (at leading and second order) and the final term contributes to $\d^2[\d_\L^{(1)}]$. Moreover, for  $\G_2 \d$ we have
\bea 
 \Shell^2_{\G_2\d}[\dlin_\L] (\vk) &=& \left[ - \frac{4}{3}\d^{(1+2)}(\vk) - \frac{376}{105}\d^{2,(2)}(\vk) +\frac{116}{105}\G_2^{(2)}(\vk) \right] \int \frac{p^2 dp}{2 \pi^2 }P_{\rm shell}(p) \vs
 && \quad \quad \quad \quad \quad \quad \quad \quad \quad \quad \quad \quad \quad \quad \quad \quad \quad \quad + \mbox{h.d.} + \O\left[\left(\dlin_\L\right)^3\right]\,.
 \label{eq:SGd}
\eea
Again, notice that (both for $\Shell^2_{\G_2\d}$ and $\Shell^2_{\d^3} $) the contributions to  $\left(\d[\d_\L^{(1)}]\right)^{(1)}$ and $\left(\d[\d_\L^{(1)}]\right)^{(2)}$ have the same prefactor. 
The Galileon operators $\G_2,\,\G_3$ and also $\Gamma_3$ will only lead to corrections to higher-derivative operators. This is a direct consequence of the non-renormalization property of the Galileon operators  \cite{Hinterbichler:2011tt,deRham:2014zqa}. Finally, higher-derivative terms such as $\nabla^2 \d$ can only source higher-derivative terms, due to the external $k^2$ momenta. 
To include one example of fourth-order operators, the correction sourced by the fourth-order $\d^4$ operator is also easy to calculate:  
\bea
\Shell^2_{\d^4} [\dlin_\L](\vk) &=&  6\int_{\vp,\vp_3} P_{\rm shell}(p) \dlin_\L(\vp_3)\dlin_\L(\vk-\vp_3) + \O\left[\left(\dlin_\L\right)^3\right] \vs 
&=& 6\left(\d^2[\d_\L^{(1)}]\right)^{(2)}(\vk)\int_{\vp} P_{\rm shell}(p)  + \O\left[\left(\dlin_\L\right)^3\right] \,,
\eea
which contributes only to $\d^2[\d_\L^{(1)}]$ at the order we work in.

Putting all the contributions together, we obtain the galaxy bias RG equations directly from Eq.~(\ref{eq:b_diffnew}):
\bea
\frac{d b_\d}{d \L} &=& -\left[\frac{68}{21}b_{\d^2}+3b_{\d^3}^\ast-\frac{4}{3}b_{\G_2\d}^\ast\right]\frac{d \s^2_\L}{d \L}\,, \label{eq:drun}
\\
\frac{d b_{\d^2}}{d \L} &=& -\left[\frac{8126}{2205}b_{\d^2} + \frac{17}{7}b_{\d^3}^\ast  - \frac{376}{105}b_{\G_2\d}^\ast + b_{n=4}^{\ast(\d^2)}  \right]\frac{d \s^2_\L}{d \L}\,, \label{eq:d2run}
\\
\frac{d b_{\G_2}}{d \L} &=& -\left[  \frac{254}{2205 }b_{\d^2} + \frac{116}{105 }b_{\G_2\d}^\ast + b_{n=4}^{\ast(\G_2)} \right]\frac{d \s^2_\L}{d \L}\,. \label{eq:Grun}
\eea
The system of equations above is the analogue of the $\beta$-function running for QFT, also known as the Callan-Symanzik equation \cite{Callan:1970yg,Symanzik:1970rt}. We use the notation $b_{n=4}^{\ast(\d^2)} = 6b_{\d^4}^\ast + \dots$ to account for the fourth order contributions to $\d^2$ and similarly for $b_{n=4}^{\ast(\G_2)}$, which, as noted above, we have not derived.
We use the superscript $b^\ast$ to indicate bias parameters that do not evolve in $\L$ and are fixed at some scale $\L_\ast$. Note that we already approximated the third and fourth-order bias parameters as constants; computing the running of those parameters would demand calculating the shell contractions for sixth-order operators. We comment on this approximation and how to go beyond it below.

We also find that the sourced higher-derivative terms have a different running compared to the other bias parameters, proportional to
 \be
\int_{\vp} \frac{P_{\rm shell}(p)}{p^2} = \int_\L^{\L+\leps} \frac{dp}{2 \pi^2 }\Plin(p) \,. \label{eq:high_der_scaling}
\ee
Higher-derivative terms only source higher-derivative terms, and hence do not affect the running of the other bias parameters (while being sourced by them, however).

\subsection{Quantitative results and closure of RG hierarchy}

We highlight that the RG equations form a linear ODE system\footnote{This is the case solely for Gaussian initial conditions. Non-Gaussian initial conditions will introduce nonlinear running of the bias and stochastic parameters \cite{Carroll:2013oxa}.} for the bias parameters in which, in general, the bias coefficients of order $n$ contribute to bias parameters at order $n-2$ and higher. 

All dependencies on cosmological parameters of the RG-flow we have derived are absorbed by $d \s^2_\L/d \L$. This is a consequence of the fact that we have assumed the Einstein-de Sitter form of the perturbation-theory kernels. Including the cosmology dependence in the kernels \cite{lssreview,Garny:2020ilv} would lead to very small, cosmology-dependent corrections to the coefficients in the RG equations. In our case however, we can change the integration variable to $\s^2$ to have cosmology-independent equations
\bea
\frac{d b_\d}{d \sigma^2} &=& -\left[\frac{68}{21}b_{\d^2}+b_{n=3}^{\ast(\d)}\right]\,, \label{eq:drun_sigma}
\\
\frac{d b_{\d^2}}{d \sigma^2} &=& -\left[\frac{8126}{2205}b_{\d^2} + b_{n=3+4}^{\ast(\d^2)}  \right]\,, \label{eq:d2run_sigma}
\\
\frac{d b_{\G_2}}{d \sigma^2} &=& -\left[  \frac{254}{2205 }b_{\d^2} + b_{n=3+4}^{\ast(\G_2)} \right]\,, \label{eq:Grun_sigma}
\eea
in which we also included the third-order bias $\{b_{\d^3}^\ast,b_{\G_3}^\ast,b_{\G_2\d}^\ast,b_{\Gamma_3}^\ast\}$ as (constant) source terms using the notation $ b_{n=3+4}^\ast =  b_{n=3}^\ast +  b_{n=4}^\ast$.
Thus, the ODE system above forms a complete set and the equation for $b_{\d^2}$ decouples from the others. Neglecting the running of higher-order bias coefficients, i.e. setting them to constants, and for initial conditions $\vb^\ast = \vb\,(\s^2=\s^2_\ast) =\left(b_\d^\ast, b_{\d^2}^\ast, b_{\G_2}^\ast \right)$, where $\sigma_*^2 \equiv \sigma_{\L_*}^2$, we find: 
\bea
b_{\delta }(\s^2 ) &=& 
 b_{\delta}^\ast + b_{\delta ^2}^\ast \frac{c_{\delta ,\delta ^2}}{c_{\delta ^2,\delta ^2}} \left(e^{-(\s^2 -\text{$\s^2_\ast $}) c_{\delta ^2,\delta ^2}}-1\right) \label{eq:dsol}\\
&&\quad-\,b_{n=3}^{\ast(\d)} (\s^2 -\text{$\s^2_\ast$})
-  b_{n=3+4}^{\ast(\d^2)}  \frac{c_{\delta ,\delta ^2}}{c_{\delta ^2,\delta ^2}^2} \left((\text{$\s^2 $}-\s^2_* ) c_{\delta ^2,\delta ^2}+e^{-(\s^2 -\text{$\s^2_* $}) c_{\delta ^2,\delta ^2}}-1\right) \,, \nonumber
\\
   b_{\delta ^2}(\s^2 ) &=&  b_{\delta ^2}^\ast  e^{-(\s^2 -\text{$\s^2_\ast $}) c_{\delta ^2,\delta ^2}} \label{eq:d2sol} \\
 && \quad +   b_{n=3+4}^{\ast(\d^2)}  \left( \frac{   e^{-(\s^2 -\text{$\s^2_\ast $}) c_{\delta ^2,\delta ^2}}}{c_{\delta ^2,\delta ^2}}  - \frac{1}{c_{\delta ^2,\delta ^2}}\right)\,, 
   \vs
b_{\G_2}(\s^2 ) &=&  b_{\G_2}^\ast + b_{\delta ^2}^\ast \frac{c_{\G_2,\delta ^2}}{c_{\delta ^2,\delta ^2}} \left(e^{-(\s^2 -\text{$\s^2_\ast $}) c_{\delta ^2,\delta ^2}}-1\right)\label{eq:Gsol} \\
&&\quad -\, b_{n=3+4}^{\ast(\G_2)}  (\s^2 -\text{$\s^2_\ast $}) 
-   b_{n=3+4}^{\ast(\d^2)}  \frac{c_{\G_2,\delta ^2}}{c_{\delta^2,\delta ^2}^2} \left((\text{$\s^2 $}-\s^2_\ast ) c_{\delta ^2,\delta ^2}+e^{-(\s^2 -\text{$\s^2_\ast $}) c_{\delta ^2,\delta ^2}}-1\right)\,,   \nonumber
\eea
with $c_{\delta,\delta ^2} = 28/61$, $c_{\delta ^2,\delta ^2} = 8126/2205$ and $c_{\G_2,\delta ^2} = 254/2205$. Notice that we are fixing the bias parameters at a reference scale $\L_\ast$. We discuss later the running of the bias for different initial conditions.

To recap, we have derived the complete set of contributions \emph{from} operators up to including third order (and $\d^4$) \emph{to} operators up to second order.  This allows us to consistently compute the running of bias coefficients up to second order, \emph{assuming} third and higher orders are zero at the initial scale $\Lambda_*$. If we include nonzero third-order operators at $\Lambda_*$, we have to make assumptions about how they evolve with $\Lambda$. We will illustrate these different cases below, and discuss how the bias RG equations are truncated. Note the difference to classic RG flow examples in renormalizable field theories, which only have a finite number of interaction vertices: in the present case, and as usual in effective field theories, new interaction vertices (bias operators) appear at each order. We split the discussion of this system of equations and solutions into three scenarios: first, by setting high-order operators to zero $b_{n\geq3}^\ast = 0$; second, considering $b_{n=3}$ and $b_{n=4}$ as constants and third, we comment on the expected behavior when taking into account the full set of dynamical ODE's at higher orders and how to truncate it. The first two scenarios are based directly on the solutions above. Later we discuss solutions for the third scenario.

\begin{figure}[ht]
    \centering
    \hspace*{0.5cm}\includegraphics[width = 0.45\textwidth]{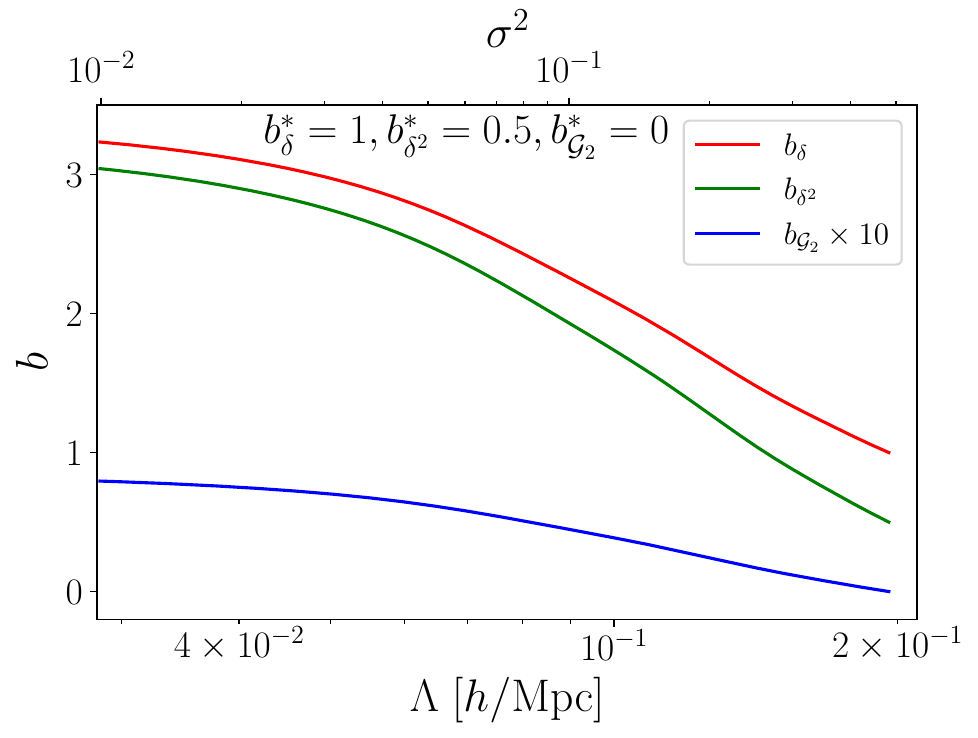}
    \includegraphics[width = 0.45\textwidth]{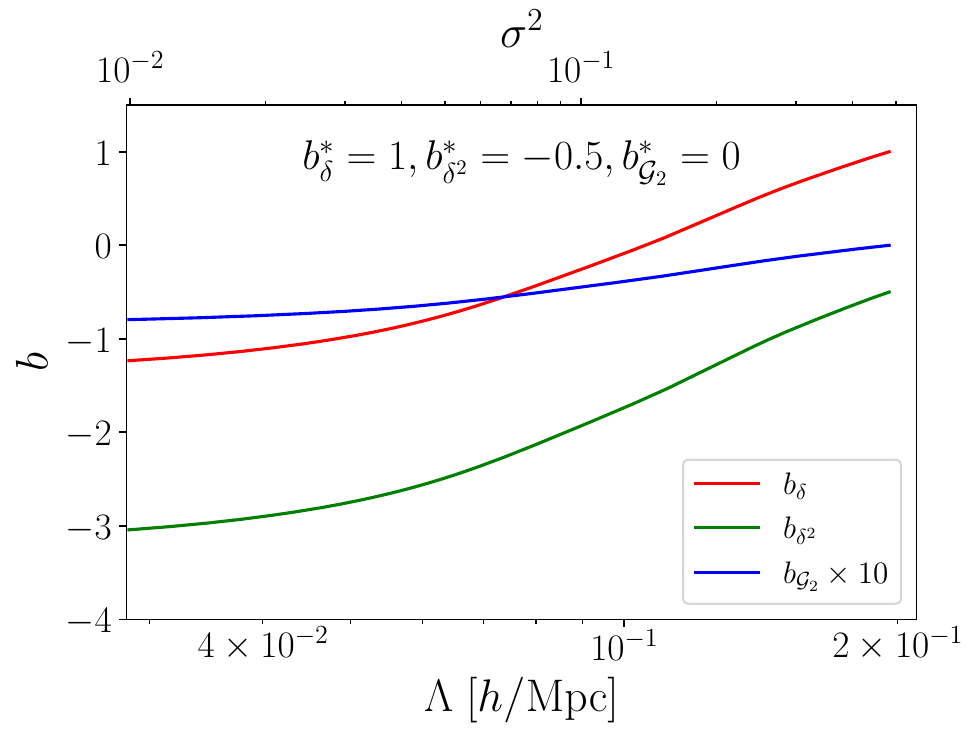}\\
    \includegraphics[width = 0.45\textwidth]{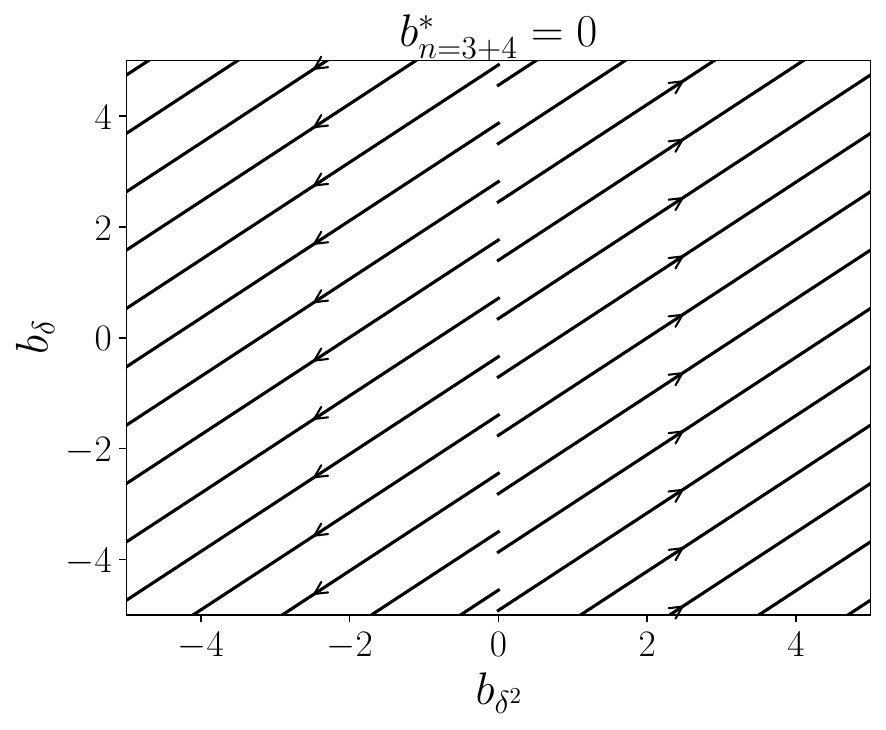}
    \hspace*{0.2cm}\includegraphics[width = 0.45\textwidth]{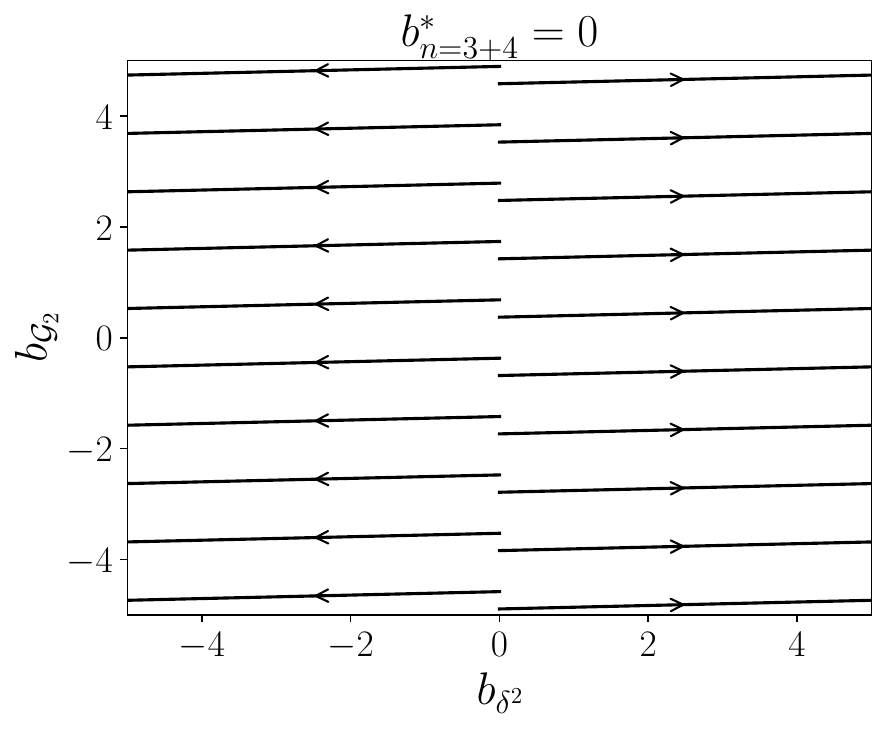}\\
    \caption{Bias RG-flow for zero higher-order bias coefficients. The top panels show the bias coefficients as a function of the cutoff $\L$ for two different initial conditions set at $\s^2_\ast = 0.5$. The bottom panels show the trajectories for different initial conditions in the $b_{\d^2},b_{\d}$ (left) and $b_{\d^2},b_{\G_2}$ (right) planes running towards $\L \to 0$.  }
    \label{fig:bias_running}
\end{figure}

\paragraph*{No higher-order bias  ($b_{n=3+4}^{(O)} = 0$).} Starting from the scenario in which higher-order operators are zero, we display in the top panels of Fig.~\ref{fig:bias_running} the bias RG-flow as a function of $\s^2$ and the cutoff $\L$ (for which we assume a $\L$CDM cosmology \cite{Planck:2018vyg}). As examples, we set two different initial conditions $\vb^\ast$ at $\s^2_\ast = 0.5$: $\left(b_\d^\ast, b_{\d^2}^\ast, b_{\G_2}^\ast \right) = \left(1, 0.5, 0 \right)$ (top left) and $\left(b_\d^\ast, b_{\d^2}^\ast, b_{\G_2}^\ast\right) = \left(1, -0.5, 0 \right)$ (top right). Note that in both cases $b_{\G_2}$ is generated by $b_\d^2$ despite setting it to zero in the initial conditions, which is in tandem with the findings of \cite{Assassi2014}. It is also worth mentioning that $b_1$ can change sign depending on the sign of $b_{\d^2}$ in this scenario. In the bottom panels of Fig.~\ref{fig:bias_running}, the lines indicate the evolution of the trajectories in the $b_{\d^2},b_{\d}$ (left) and $b_{\d^2},b_{\G_2}$ (right) planes when running towards $\L \to 0$ starting from different initial conditions. Note that the running of $b_{\G_2}$ is substantially slower compared to the other bias parameters. The stability analysis of the system indicates that the only (unstable) fixed point is at $b_{\d^2} = 0$. We see from the top panel of Fig.~\ref{fig:bias_running} that the bias parameters exponentially depend on the cutoff but asymptote to a finite constant on large scales, determining the endpoints of the lines seen in the bottom panels.\footnote{As an aside, note from Eqs.~(\ref{eq:dsol})-(\ref{eq:Gsol}) that, for the (non-perturbative) limit $\s^2 \to \infty$ we find that $b_{\delta ^2}\to 0$ and $b_{\d},b_{\G_2 }$  approach a constant. In this limit, however, second-order operators also source higher-order operators that back-react and change this solution.} 

\begin{figure}[ht]
    \centering
    \hspace*{0.5cm}\includegraphics[width = 0.45\textwidth]{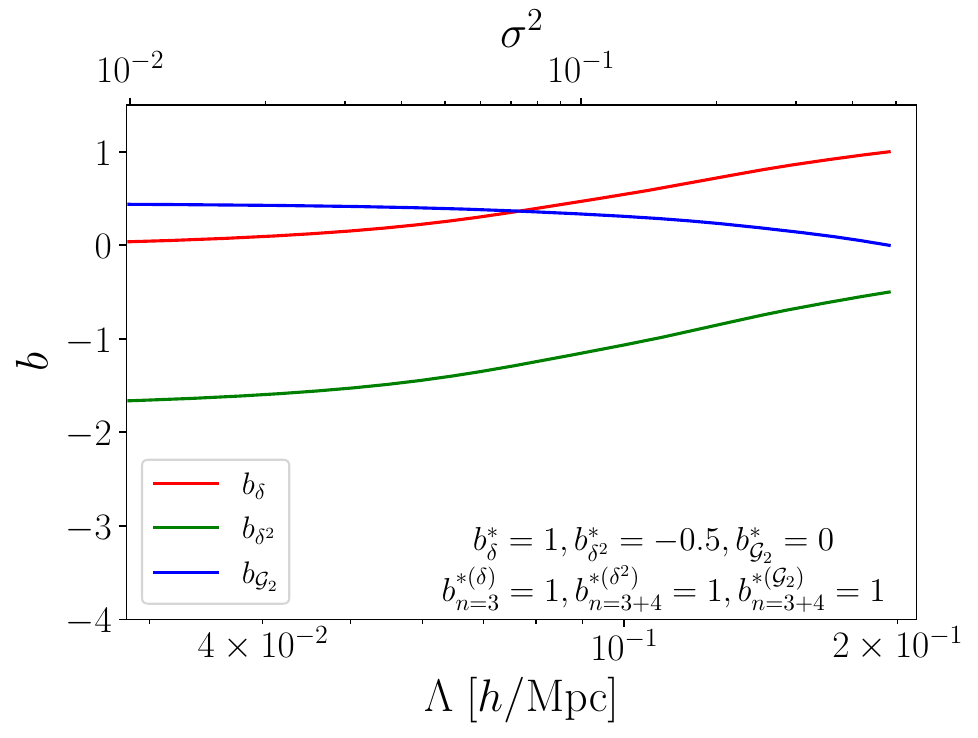}
    \includegraphics[width = 0.45\textwidth]{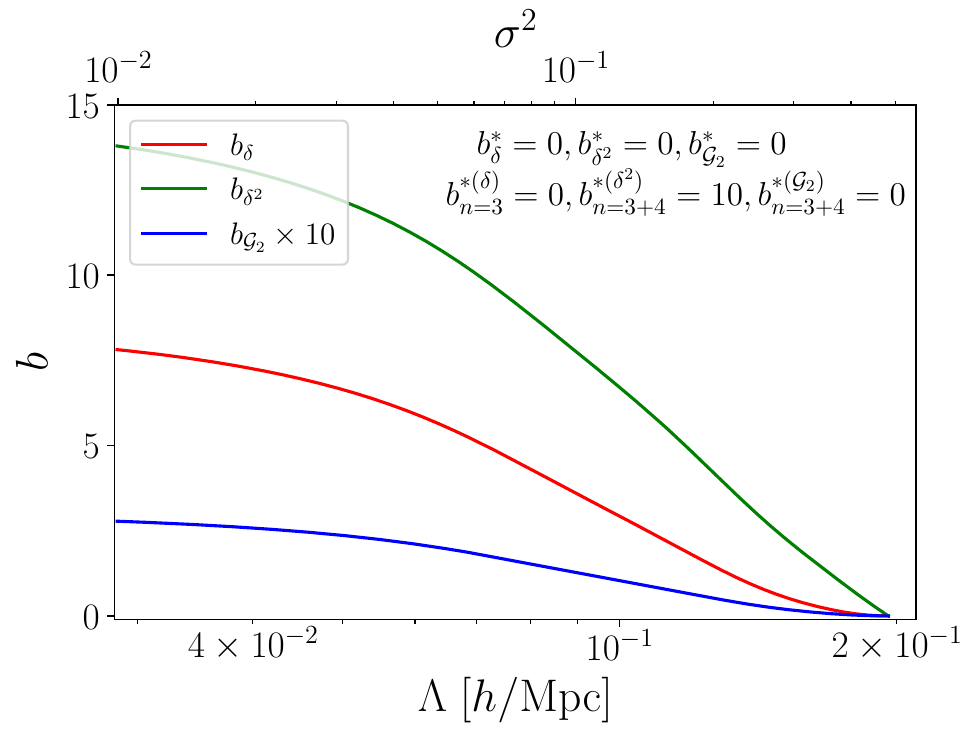}\\
    \includegraphics[width = 0.45\textwidth]{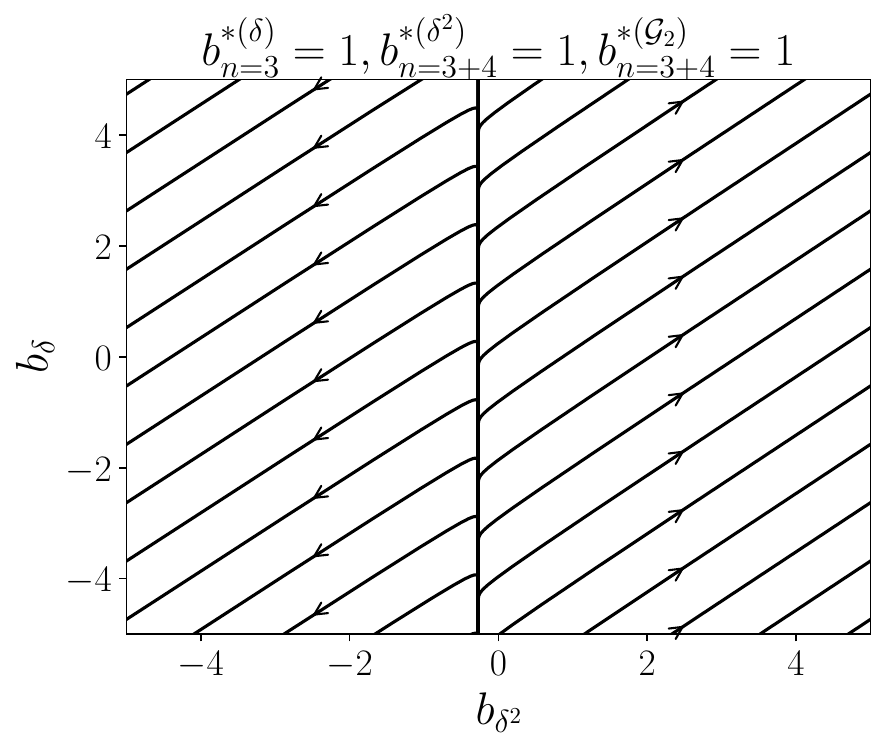}
    \hspace*{0.2cm}\includegraphics[width = 0.45\textwidth]{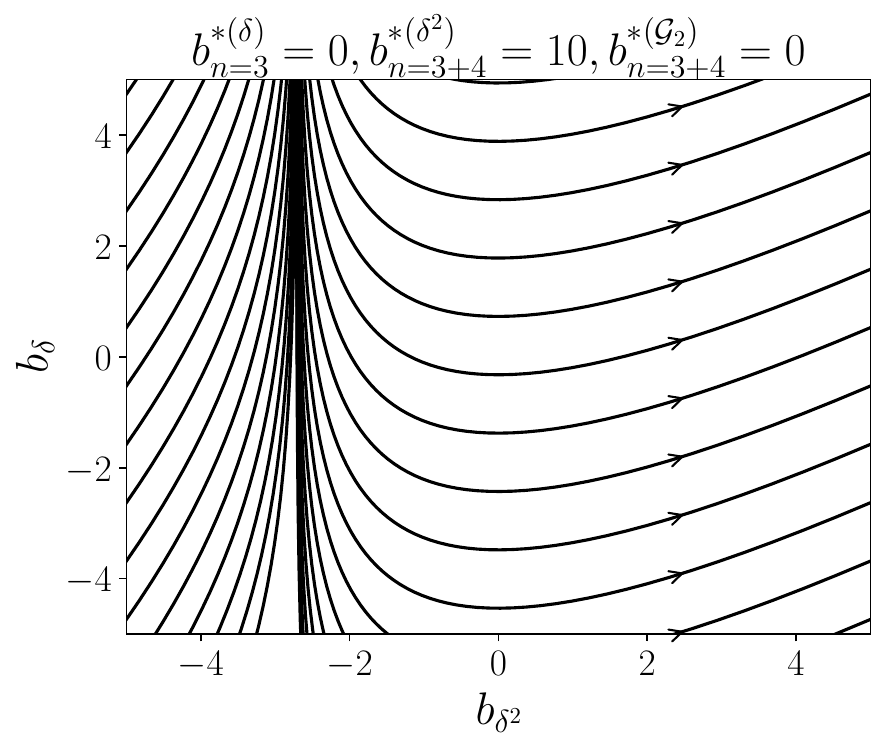}\\
    \caption{Same as \reffig{bias_running}, but with higher-order bias parameters as constant sources. The left and right panels show two different initial conditions set at $\s^2_{*} = 0.5$. 
    }
    \label{fig:bias_running2}
\end{figure}

\begin{figure}[ht]
    \centering
    \includegraphics[width = 0.45\textwidth]{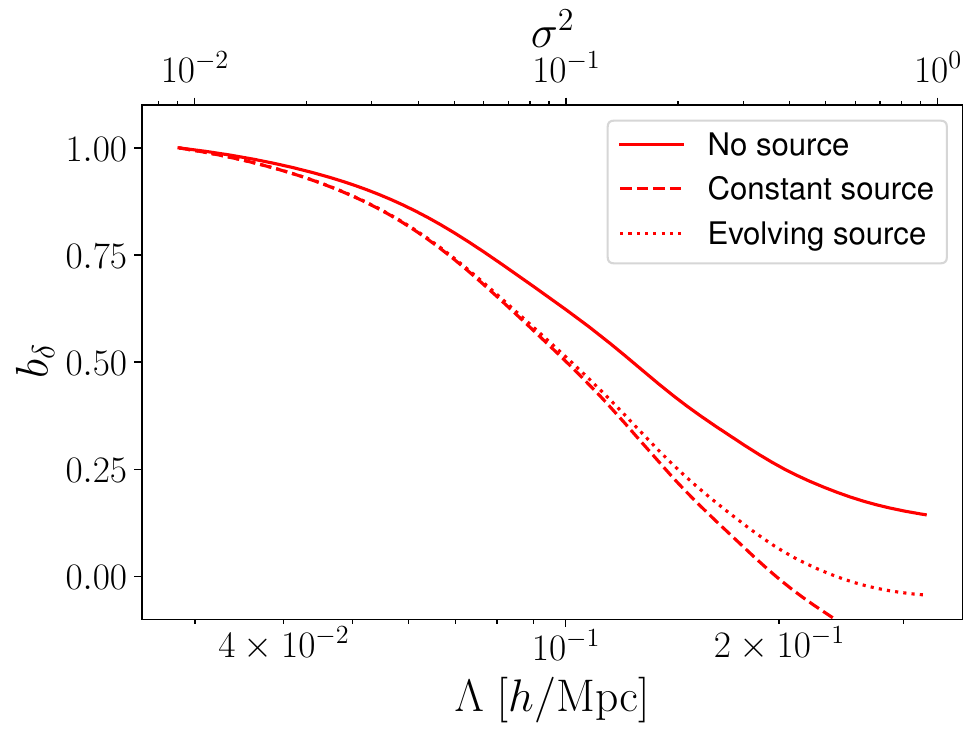}
    \includegraphics[width = 0.45\textwidth]{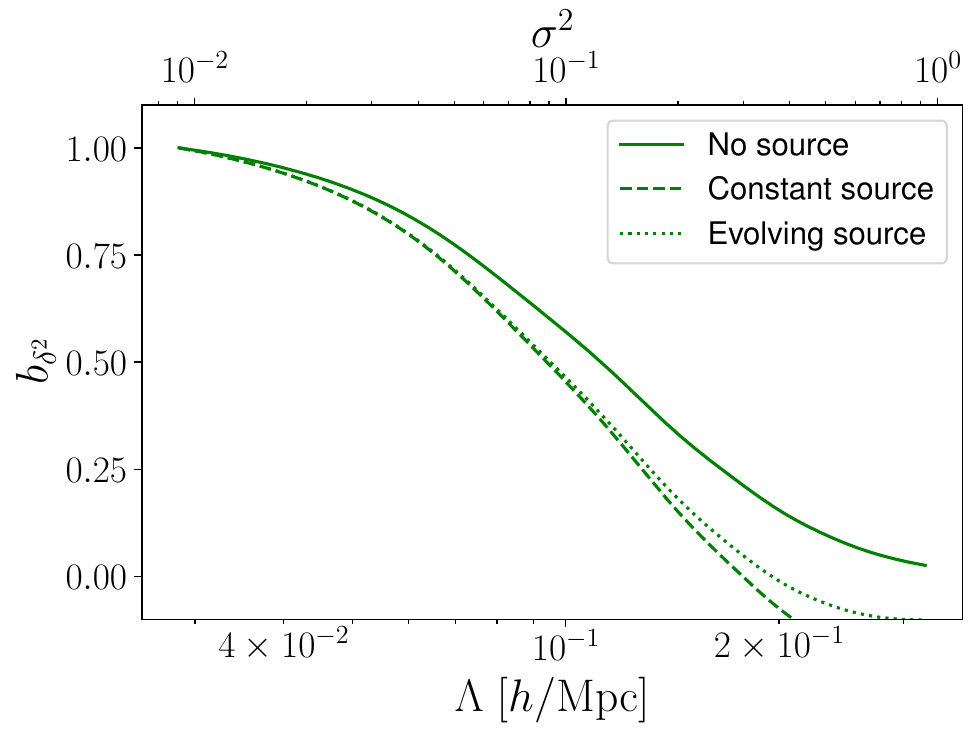}\\
    \includegraphics[width = 0.45\textwidth]{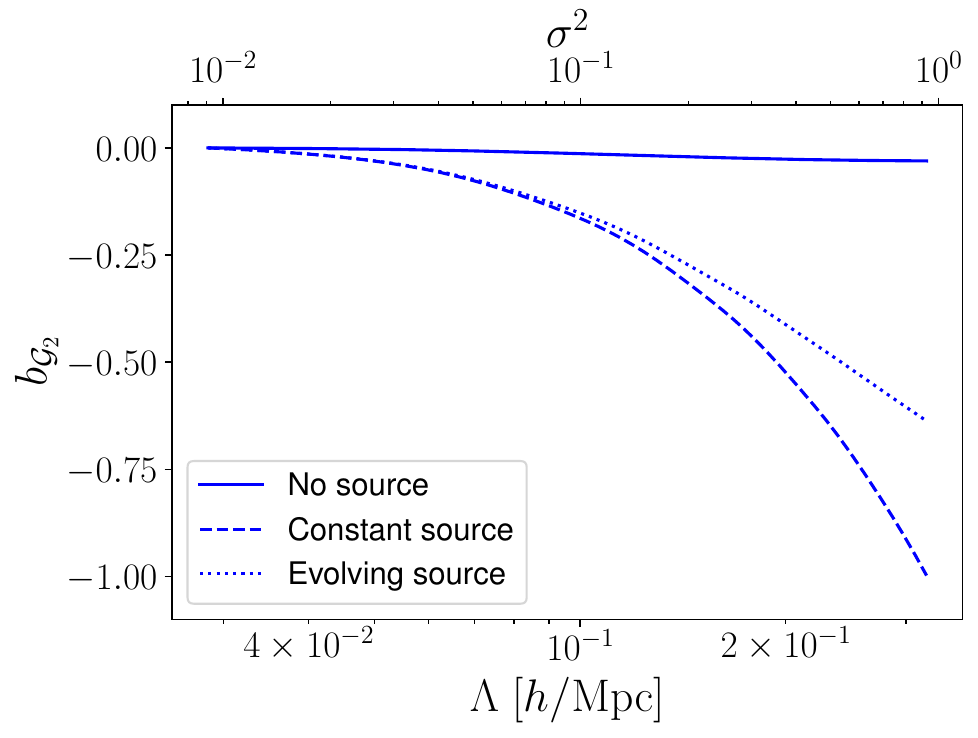}
    \caption{Comparison of the RG-running for $b_\d$ (top left), $b_{\d^2}$ (top right) and $b_{\G_2}$ (bottom) starting from $\L_* = 0$ ($\s^2_\ast = 0$) to higher $\L$. The solid lines show the result in the absence of higher-order bias, while the dashed and dotted lines show the result for constant and running, via Eq.~(\ref{eq:bansatz}), higher-order bias parameters. We have fixed $\left(b_\d^{\ast}, b_{\d^2}^{\ast}, b_{\G_2}^{\ast}\right) = \left(1, 1, 0 \right)$ as initial conditions,  $b_{n=3}^{\ast(\d)} = b_{n=3+4}^{\ast(\d^2)} = b_{n=3+4}^{\ast(\G_2)} = 1$ (for the dashed and dotted cases) and $c^{(\d)} = c^{(\d^2)} = c^{(\G_2)} = 1$ (dotted case). We see that, at sufficiently low $\Lambda$, the running of the higher-order coefficients becomes irrelevant, showing that the RG equations can be truncated. }
    \label{fig:scenarios_ratio}
\end{figure}

\paragraph*{Constant higher-order bias   ($b_{n=3+4}^{(O)} = b_{n=3+4}^{\ast(O)}$).} In the second scenario, we include third and fourth-order terms as static sources. The top panels of Fig.~\ref{fig:bias_running2} show the running of the bias parameters for fixed initial conditions at $\s^2_\ast = 0.5$. The left panel represents the scenario in which all higher-order parameters are order one. For that case, we still see the sign change for the bias parameters (especially dependent on the sign of $b_{\d^2}$). In the right panel, we take $b_{n=3+4}^{\ast(\d^2)} = 10 \gg 1$ as the only non-vanishing parameter. In that case, the other bias parameters are generated by this source term. The bottom panels indicate the trajectories in the $b_{\d^2},b_{\d}$ plane for those two different initial conditions for higher-order terms. When the source becomes parametrically large compared to the bias parameters (right), more complicated trajectories start to form. 
In the case the sources are order one (left panel), the system behaves similarly to the unsourced scenario, but we can notice significant quantitative changes.  This is shown in Fig.~\ref{fig:scenarios_ratio}, where we compare the bias evolution without (solid lines) and with constant higher-order bias parameters (dashed lines). There is a noticeable difference between both scenarios even at low $\Lambda$, showing that bias terms of order $n+2$ cannot be neglected in the calculation of the running of $O^{[n]}$.

\paragraph*{Running higher-order bias and truncation of RG hierarchy.} Finally, we discuss the truncation of the hierarchy of bias RG equations. As mentioned before, the complete running of the second-order bias parameters demands calculating the fourth-order bias terms and those, in turn, also depend on the running of sixth-order bias parameters. One may then wonder if there is a systematic way to truncate this ODE system. Here we argue that, in order to calculate the complete evolution of operators of order $n$, it is sufficient to truncate the series at order $n+2$, at least if one is interested in scales much below the initial scale $\L_*$. This means that for the calculation of the RG-flow for second-order operators it is sufficient to consider third and fourth-order operators as static sources (i.e. not including their RG equations which involve bias parameters that are of fifth and higher order). 
Proving that this holds at all orders is beyond the scope of this paper, but since the bias expansion guarantees that (at least for large scales) the system is perturbative, we take the heuristic approach of considering
\bea \label{eq:bansatz}
 b_{n=3+4}^{(O)}(\s^2) = b^{*(O)}_{n=3+4}\, e^{-c^{(O)}(\s^2 - \s^2_\ast)}  
\eea
as an ansatz for the higher-order bias evolution, resembling the solutions found in Eqs.~(\ref{eq:dsol})-(\ref{eq:Gsol}). 
The two free parameters $b^{*(O)}$ and $c^{(O)}$  for each operator $O$ are, respectively, a proxy for the higher-order bias coefficients and the shell prefactors (analogous to $c_{\delta,\delta ^2},\, c_{\delta ^2,\delta ^2}$ and $c_{\G_2,\delta ^2}$). 
Notice that we do not expect a hierarchy in these coefficients [see Eqs.~(\ref{eq:d2run})-(\ref{eq:Grun})].
After considering an ansatz of the type~(\ref{eq:bansatz}) for $b_{n=3}^{(\d)}(\s^2)$, $b_{n=3+4}^{(\d^2)}(\s^2)$ and $b_{n=3+4}^{(\G_2)}(\s^2)$ we can find an analytical solution for the running (see Appendix~\ref{eq:proxisol}), which we show as dotted lines in Fig.~\ref{fig:scenarios_ratio}. We find that the running of the third and fourth-order bias parameters do not affect the running of second-order parameters  for sufficiently large scales, as long as their coefficients $b^{(O)}$ and $c^{(O)}$ are $\O(1)$.

To calculate the $\L$ evolution of the second-order parameters it is therefore sufficient to truncate the ODE system at second order, and to include third-order and fourth-order terms as static sources. The parametrized running described here could also be included as an estimate of the theory error in the running.

\subsection{Main conclusions}
We now summarize the relevant conclusions from the bias running:
\begin{itemize}
    \item In general, an operator $O$ of order $n$ will contribute to operators of order $n-2$ and higher, which is a direct consequence of the shell contractions in $\Shell^2_O$. 
      Therefore, the calculation up to third-order terms presented here gives the complete running of $b_\d$. In order to obtain the full running of second-order operators, one would need to calculate fourth-order corrections, which we leave for future work. Also note that an $n$-th order operator in general sources operators that are higher than $n$-th order, for example:
  \be
  \raisebox{0.22cm}{
    \includegraphicsbox[scale=1]{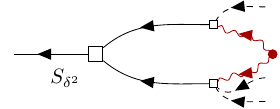}}
  \raisebox{-0.0cm}{
    \includegraphicsbox[scale=1]{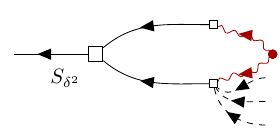}} \qquad.
  \label{eq:d2tod3}
  \ee
  The ODE system is then vertically coupled: even though the system may start with zero higher-order operators, those will be sourced and backreact affecting the running of lower-order terms. This effect is, however, perturbatively suppressed.
    \item In spite of the RG-flow equations being coupled at all orders, we find strong indications, though not a rigorous proof, that the relevant contributions for the running of an operator of order $n$ come from operators that are up to order $n+2$ (\reffig{scenarios_ratio}). Higher-order operators only indirectly couple to the lower-order terms, and are suppressed on large scales (i.e., low $\L$). This points to a natural way to truncate the ODE system. 
    \item Individual contributions from one term in the shell integrals can source operators that break Galilean invariance, see Eqs.~(\ref{eq:shell_d2}), (\ref{eq:F2F2}) and (\ref{eq:F3}). When summed, those contributions cancel out.
    Similar findings were reported for the $\d^2$ renormalization in \cite{Assassi2014}.
    \item In the expansion of $\Shell^2_{O'}$ in operators $O$, the contribution to an operator $O^{(n)}$ is the same for different perturbative orders $n$
    [see for example the prefactors of $\d$ in Eqs.~(\ref{eq:Sd2}), (\ref{eq:Sd3}) and (\ref{eq:SGd})]. This fact can be understood as a statement that the bias parameters $b_O^\L$ are physical and independent of the order in perturbation theory that the running is calculated at.
    Ultimately, this also is a consequence of Galilean symmetry, and implies that the Galilei- or boost-invariant basis of operators \cite{MSZ,senatore:2014} remains complete. The RG flow could correspondingly be used to check for completeness of the bias expansion at higher orders.
    \item Contractions of two lines going into the same $F_\ell$ kernel,
    represented by the top right diagram in Fig.~\ref{fig:Sdiag}, source higher-derivative operators.
    \item Shell contractions between two different bias operators source higher-order and higher-derivative stochastic contributions. Therefore, a complete description of the galaxy density field has to include those terms. 
\end{itemize}
In practice, one would vary the scale $\L$, either at fixed $k_{\rm max}$ or keeping $k_{\rm max}$ as a fixed fraction of $\L$, to confirm that parameters that should be $\L$-independent, such as cosmological parameters, are in fact so. One can also compare the inferred running of bias coefficients with that predicted by the theory as derived here as a check of the relevance of higher-order contributions. In any case, we expect the perturbative treatment to break down once $\L$ approaches $k_{\rm NL}$. 

\section{Connection to bias parameters defined via \texorpdfstring{$\bm{n}$}{n}-point functions} \label{sec:connection}

The (so far) standard approach to comparing theory with data is to use $n$-point correlation functions, where the theory prediction is usually calculated semi-analytically as the ensemble mean of a given correlation function in perturbation theory. These correlation functions have leading and next-to-leading order contributions; the latter in general need to be regularized. In this context, Ref.~\cite{Assassi2014} derived renormalization conditions [\refeq{renormcond_pre}] which determine what counterterms need to be subtracted from a given bias operator in order to regularize loop integrals [e.g., \refeq{renorm_d2}]. For this, they introduced a cutoff, but merely as a calculational tool; in practice, this cutoff is usually sent to infinity when numerically evaluating loop integrals, as they converge to finite values at low loop orders (at least in $\Lambda$CDM). Note the key difference to the cutoff used so far in this paper, which is kept finite and explicit throughout.
According to the Wilsonian prescription, and as we comment later, the observables (i.e. the $n$-point functions) in the end have to be independent of this artificial (finite or infinite) cutoff while the \emph{free coefficients} (coupling constants and masses, or bias parameters) in general depend on $\L$. The information absorbed in the bias coefficients is different in both approaches, however, since the finite cutoff allows us to keep more perturbative modes in the loop integrals which are removed in the standard renormalized bias scheme.

In this section, we derive the connection between the \emph{finite-scale} operators, where $O = O[\dlin_\L]$ is constructed from the filtered density field at the scale $\L$,\footnote{We emphasize once more that $O = O[\dlin_\L]$ is a short notation for $O = O[\d[\dlin_\L]]$.} and the \emph{renormalized} bias operators $\renormOP{O}$, following \cite{Assassi2014} (see also Sec.~2.10 of \cite{Desjacques:2016bnm}). The connection between both frameworks is done in the ``infrared'' limit $\L \to 0$; precisely, it is given by the relation
\be
\lim_{\L \to 0;\  k/\Lambda\  {\rm fixed}} \< O[\dlin_\L](\vk) \renormOP{O'}(\vk') \> =  \lim_{\L \to 0;\  k/\Lambda\  {\rm fixed}} \< O[\dlin_\L](\vk) O'[\dlin_\L](\vk') \> \,.
\label{eq:antecipate}
\ee
The goal of this section is to prove this relation. The reason we look at the infrared limit is that first, this is what the definition of renormalized operators refers to, and second, the calculation becomes much more tractable and transparent in this limit. Readers who mainly want to understand the relevance of the choice between finite-scale and $n$-point renormalization scheme for observables such as the galaxy power spectrum may skip directly to \refsec{diffcontrib}.

The renormalized operators are defined via renormalization conditions at the $n$-point correlator level; specifically, one demands that for all $m\geq 0$,
\be
\langle \d^{(1)}(\vk_1)\cdots\d^{(1)}(\vk_m) \renormOP{O}(\vk) \rangle
\stackrel{k_i \to 0}{\longrightarrow}
\langle \d^{(1)}(\vk_1)\cdots\d^{(1)}(\vk_m) O[\d](\vk) \rangle_{\rm LO} ,
\label{eq:renormcond}
\ee
where the subscript LO denotes the leading-order contribution in perturbation theory, which we will define below.
To provide some context for this definition, note that any $n$-th order renormalized bias operator can be unambiguously defined using the tree-level $(n+1)$-point function on large scales. For example, $\renormOP{\d}$ can be defined by the tree level power spectrum in the limit $k\to 0$, and similarly for $\renormOP{\d^2}$ and $\renormOP{\G_2}$ via the large-scale bispectrum. The renormalization condition \refeq{renormcond} only becomes nontrivial when loop contributions are considered. They are enforced via counterterms to the renormalized operators which ensure that the bias parameters measured do not depend on small-scale non-perturbative modes.
In summary, \refeq{renormcond} implies that $\renormOP{O}$ is constructed following \refeq{Odef}, but with additional counterterms at a given order in perturbations [see \refeq{renorm_d2} for an example]. We will not need the explicit expressions of the counterterms to obtain the connection to the field-level operators, but comment on them later in the section.

Our goal in this section is to show that the bias coefficients $b_O^\Lambda$ agree with those measured using $n$-point functions, $b_{O}^{\rm n-pt}$ on large scales, i.e. in the limit of $\Lambda$ and $k$ going to zero.
Apart from the conceptual connection between both schemes, this also allows for a practical conversion of bias parameters. Specifically, one would measure $b_O^{\Lambda_\ast}$ at some scale $\Lambda_\ast$, and then run these down to $\Lambda\to 0$ via the running derived in the previous section. The resulting bias coefficients are then equivalent to the coefficients $b_O^{\rm n-pt}$ of the renormalized operators.

The equivalence of $b_O^\Lambda$ and $b_{O}^{\rm n-pt}$ in the infrared limit can be made precise by stating that the corresponding operators agree in a certain limit: 
\be
L_{OO'} \equiv \lim_{\L \to 0;\  k/\Lambda\  {\rm fixed}} \< O[\dlin_\L](\vk) \renormOP{O'}(\vk') \> =  R_{OO'} \equiv \lim_{\L \to 0;\  k/\Lambda\  {\rm fixed}} \< O[\dlin_\L](\vk) O'[\dlin_\L](\vk') \> .
\label{eq:toshow}
\ee
Notice that we take the large-scale limit keeping the ratio $k/\Lambda < 1$ fixed. This ensures that $O[\dlin_\L]$ is only evaluated at $k < \L$ while we lower the cutoff. 
Suppose now that \refeq{toshow} holds. 
Since both the set $\{\renormOP{O}\}$ and $\{O[\dlin_\L]\}$ form a complete basis at any given order (and for any finite $\L$), we can write for a given tracer $\d_g$
\be
\d_g = \sum_O b_O^\L \,O[\dlin_\L] = \sum_{O'} b_{O'}^{\rm n-pt} \,\renormOP{O'},
\ee
where both sums run over the same complete set of bias operators 
up to some arbitrary but fixed order. Multiplying both sides by $O''[\dlin_\L]$ and taking the ensemble average,
we see that
\be
M_{OO'} \equiv \sum_{O''} (R^{-1})_{OO''} L_{O'O''}
\ee
is the invertible transformation matrix between the two operator conventions $\renormOP{O}$ and $O[\dlin_\L]$. The bias coefficients correspondingly transform with $(M^{-1})^\top$. 
\refeq{toshow} thus says that in the $\L\to 0$ limit, the two bases agree, i.e. $M_{OO'}$ is the identity, and hence the bias coefficients $b_O$ likewise agree for any given tracer.

Before delving into the derivation of \refeq{toshow},
it is useful to make the following distinction for the contributions to such correlators: 
\begin{itemize}
\item \emph{Gaussian:} If instances of $\d(\vp_i)$ from $O$ and $O'$ are set to $\d^{(1)}(\vp_i)$ when inserting \refeq{Odef} into the operator correlators in \refeq{toshow}, we call this a ``Gaussian contribution.'' When correlating an $n$-th order operator with an $m$-th order one, this involves integration over precisely $n+m$ momenta. Consequently, after removing Gaussian counterterms [see \refeq{gaussiancounter}], this contribution vanishes if $n\neq m$. If $n = m$, this contribution is of order $(\Plin)^n$. We mostly single out the Gaussian contributions as a ``warmup'' to the general case.
\item \emph{Irreducible:} Consider the cross-correlation of two operators evaluated at the same perturbative order $n$, $\< O^{(n)}(\vk) O'^{(n)}(\vk')\>$. 
We refer to \emph{$n$-particle-irreducible diagrams} contributing to this correlator (or simply ``irreducible contributions'') as those for which all $n$ linear propagators are connected between $O$ and $O'$. The name follows from the fact that all $n$ linear propagators would need to be cut to separate the diagram into two different diagrams.
Note that, for any $O,O'$ these contributions exist at arbitrarily high orders in perturbation theory. We will show that irreducible contributions on both sides of \refeq{toshow} agree.
\item \emph{Reducible:} Consider cross-correlation of two operators $\< O^{(n)}(\vk) O'^{(m)}(\vk')\>$ of in general different perturbative orders. This correlation involves $(n+m)/2$ linear propagators and vanishes if $n+m$ is odd, as we consider Gaussian initial conditions. We refer to $(n+m)/2$-particle-reducible contributions (or simply \emph{reducible contribution}) as those for which \emph{not} all linear propagators are connected between $O$ and $O'$. Note that if $n\neq m$, all contributions to the correlator are reducible. We will show that reducible contributions on both sides of \refeq{toshow} either vanish or are suppressed in the limit $\L\to 0$.
\end{itemize}
We will consider each class of contributions separately in the discussion below.

\subsection{Gaussian contribution}

\begin{figure}[t]
  \centering
    \raisebox{0.51cm}{\includegraphics[width = 0.49\textwidth]{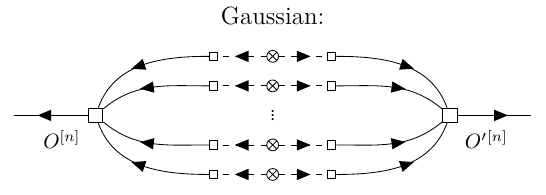}}
    \includegraphics[width = 0.49\textwidth]{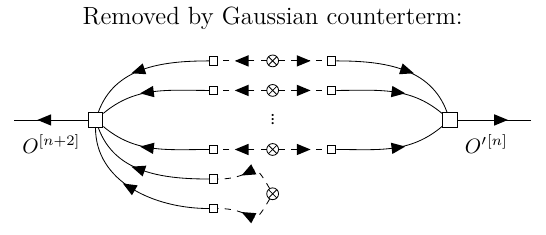}
    \caption{Example Gaussian contributions to $L_{OO'}$. The contribution on the right is absorbed by a Gaussian counterterm (e.g., $\d^2 \to \d^2 - \<\d^2\>$ and $\d^3 \to \d^3 - 3\<\d^2\> \d$). After this subtraction, as argued in the text, the Gaussian contributions agree with the corresponding contributions to $R_{OO'}$, thanks to \refeq{Pkcut}.}
    \label{fig:diags_gauss}
\end{figure}

Let us begin with the Gaussian result for $R_{OO'}$,
\ba
&R_{OO'}\Big|_{\rm Gaussian} \stackrel{\rm }{=} \lim_{\L \to 0; k/\Lambda\  {\rm fixed}} \< O[\dlin_\L](\vk) O'[\dlin_\L](\vk') \>_{\rm Gaussian} \vs
&= \lim_{\L \to 0; k/\Lambda\  {\rm fixed}} \int_{\vp_1,\ldots,\vp_n} \dirac(\vk-\vp_{1\ldots n}) S_O(\vp_1,\ldots, \vp_n)
\int_{\vp'_1,\ldots,\vp'_m} \dirac(\vk'-\vp'_{1\ldots m}) S_{O'}(\vp'_1,\ldots, \vp_m') \vs
&\qquad\qquad\times \left\< \d_\L^{(1)}(\vp_1) \ldots \d_\L^{(1)}(\vp_n) \d_\L^{(1)}(\vp'_1) \ldots \d_\L^{(1)}(\vp'_m) \right\>.
\label{eq:ROOLO}
\ea
We now require that \emph{only contractions among the $\vp_i$ and $\vp'_j$ contribute to this correlator}. This can be ensured at the operator level by subtracting what we call Gaussian counterterms. For the first few Eulerian operators, these are
\be
\d^2(\vx) \to \d^2(\vx) - \<\d^2\>; \quad
\d^3(\vx) \to \d^3(\vx) - 3\<\d^2\> \d(\vx); \quad
\mathcal{G}_2 \d(\vx) \to \mathcal{G}_2 \d(\vx) +\frac43 \<\d^2\> \d(\vx),
\label{eq:gaussiancounter}
\ee
and so on (see top panels of Fig.~\ref{fig:diags_gauss}), where all instances of $\d$ precisely denote $\d[\dlin_\L](\vx)$.
These are in fact the same contributions that appear when performing a bias expansion in Lagrangian space using orthogonal polynomials  \cite{desjacques:2013,PBSpaper,lazeyras/musso/desjacques:2015}. Since they can be calculated based completely on the assumption of Gaussian statistics of $\d$, we refer to these as \emph{Gaussian counterterms.} 
  Notably, the Gaussian counterterms were also included in the halo bias measurements of \cite{lazeyras/schmidt:2018,abidi/baldauf:2018}, who correlated operators constructed at the field level with the halo density field in simulations. 
  Notice that we have not subtracted the Gaussian counterterms in the calculation of the RG equations in the previous section. Indeed, one can check from \refeqs{Sd3}{SGd} that including these counterterms removes the lowest-order contributions there [as noted after \refeq{Sd2}, we did not write the tadpole contribution for $\d^2$].

Then, after removing the Gaussian counterterms, \refeq{ROOLO} vanishes if $m\neq n$. A nonzero contribution arises from expanding $\d_\L$ beyond linear order, i.e. inserting $F_\ell$ kernels with corresponding additional integrated momenta into \refeq{ROOLO}. We will consider this contribution below.

Now we can turn to the corresponding result for $L_{OO'}$,
\ba
& L_{OO'}\Big|_{\rm Gaussian} = \lim_{\L \to 0;\  k/\Lambda\  {\rm fixed}} \< O[\dlin_\L](\vk) \renormOP{O'}(\vk') \>_{\rm Gaussian} \vs
&= \lim_{\L \to 0;\  k/\Lambda\  {\rm fixed}} \int_{\vp_1,\ldots,\vp_n} \dirac(\vk-\vp_{1\ldots n}) S_O(\vp_1,\ldots, \vp_n) \< \d_\L^{(1)}(\vp_1) \ldots \d_\L^{(1)}(\vp_n) \renormOP{O'}(\vk') \> \vs
&= \lim_{\L \to 0;\  k/\Lambda\  {\rm fixed}} \int_{\vp_1,\ldots,\vp_n} \dirac(\vk-\vp_{1\ldots n}) S_O(\vp_1,\ldots, \vp_n) \< \d_\L^{(1)}(\vp_1) \ldots \d_\L^{(1)}(\vp_n) O'[\delta](\vk') \>_{\rm LO}. \label{eq:LOOLO}
\ea
The final equality follows from the definition of the $n$-pt renormalized operators in the large-scale limit, \refeq{renormcond}. In particular, this statement implies that there are only contractions across the $\vp_i$ and the momenta hidden in $O'[\d]$. Thus, if $n = m$, these are the same contractions as in \refeq{ROOLO}.

It then follows that the only difference between \refeq{ROOLO} and \refeq{LOOLO} are the filters involved: in \refeq{ROOLO}, both $\vp_i$ and $\vp'_j$ are cut at $\L$, while in \refeq{LOOLO} only the $\vp_i$ are cut at $\L$. However, since we have argued that only cross-contractions between $\vp_i$ and $\vp'_j$ contribute, and since
\be
\< \d_\L^{(1)}(\vp) \d_\L^{(1)}(\vp') \> = \< \d_\L^{(1)}(\vp) \d^{(1)}(\vp') \>
\label{eq:Pkcut}
\ee
for sharp-$k$ filters, it follows that \emph{\refeq{LOOLO} equals \refeq{ROOLO} for the Gaussian contributions.}

\subsection{Irreducible contributions}
\label{sec:irreducible}

\begin{figure}[t]
  \centering
    \raisebox{0.51cm}{\includegraphics[width = 0.49\textwidth]{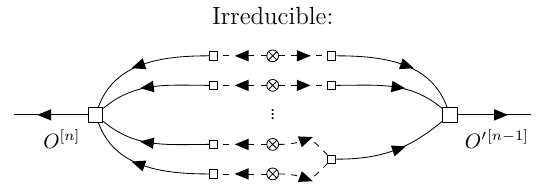}}
    \includegraphics[width = 0.49\textwidth]{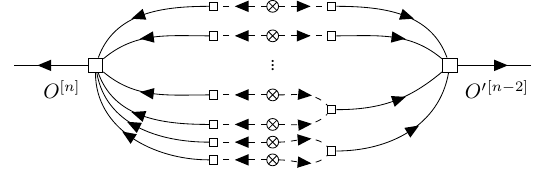}
    \caption{
      Example irreducible contributions to $L_{OO'}$. As argued in the text, they agree with the corresponding contributions to $R_{OO'}$, thanks to \refeq{Pkcut}.
    }
    \label{fig:diags_irred}
\end{figure}

Consider a contribution to $R_{OO'}$ that involves the cross-correlation of two operators evaluated at the same order $n$,
\be
R_{OO'} \supset \lim_{\L \to 0;\  k/\Lambda\  {\rm fixed}} \< O^{(n)}[\dlin_\L](\vk) O'^{(n)}[\dlin_\L](\vk')\> .
\ee
Expanding this correlator into Feynman diagrams, we can identify $n$-particle-irreducible diagrams, and $n$-particle-reducible diagrams: the latter can be separated into two separate diagrams by cutting less than $n$ linear propagators, while the former cannot. Here, we only consider the $n$-particle-irreducible diagrams, which we refer to as ``irreducible contributions'' for short. Examples are shown in \reffig{diags_irred}. It is clear that in these diagrams, each linear propagator is connected to both $O$ and $O'$. In fact, this holds for all irreducible contributions. Note that we have not specified the perturbative order $n$ here; in fact, irreducible contributions exist at arbitrarily high order.

As an example, consider $O^{(2)} = \d^{(2)}, O'^{(2)} = (\d^2)^{(2)}$. Then,
dropping the tadpole counterterm $\<\d^2\>$,
\ba
R_{\d \d^2}\Big|_{\rm irred.} =& \lim_{\L \to 0;\  k/\Lambda\  {\rm fixed}} \< \d^{(2)}[\dlin_\L](\vk) (\d^2[\dlin_\L])(\vk') \> \vs
=& \,(2\pi)^3\dirac(\vk+\vk') \lim_{\L \to 0;\  k/\Lambda\  {\rm fixed}} 2 \int_{\vp} F_2(\vp, \vk-\vp)
\Plin^\L(\vp) \Plin^\L(\vk-\vp) . 
\label{eq:Rdd2}
\ea
Let us then consider the corresponding contribution to $L_{\d\d^2}$,
\ba
L_{\d \d^2}\Big|_{\rm irred.} =& \lim_{\L \to 0;\  k/\Lambda\  {\rm fixed}} \< \d^{(2)}[\dlin_\L](\vk) \renormOP{\d^2}(\vk') \> \vs
=& \lim_{\L \to 0;\  k/\Lambda\  {\rm fixed}}
\int_{\vp} F_2(\vp, \vk-\vp) \< \d_\L^{(1)}(\vp) \d_\L^{(1)}(\vk-\vp) \renormOP{\d^2}(\vk') \vs
=& \lim_{\L \to 0;\  k/\Lambda\  {\rm fixed}}
\int_{\vp} F_2(\vp, \vk-\vp) \< \d_\L^{(1)}(\vp) \d_\L^{(1)}(\vk-\vp) (\d^2)(\vk') \>_{\rm LO} , 
\label{eq:Ldd2}
\ea
where we have used the renormalization conditions in the second line (note that we are taking both $k$ and $\L$ to zero). The non-vanishing LO contribution evaluates to the same as \refeq{Rdd2}, thanks again to \refeq{Pkcut}.
Note that $\< \d^{(1)}(\vk) (\d^2)^{(3)}(\vk')\>$ is not a \emph{reducible} contribution; we will turn to those below.

The opposite case,  $O=(\d^2)^{(2)}, O'=\d^{(2)}$, is simpler and yields
\be
L_{\d^2 \d}\Big|_{\rm irred.} = \lim_{\L \to 0;\  k/\Lambda\  {\rm fixed}} \< \d^2[\dlin_\L](\vk) \renormOP{\d}(\vk') \> = \lim_{\L \to 0;\  k/\Lambda\  {\rm fixed}} \< \d^2[\dlin_\L](\vk) \d^{(2)}(\vk') \>_{\rm LO},
\label{eq:Ld2d}
\ee
which is immediately seen to be equal to \refeq{Rdd2}.

The diagrams in \reffig{diags_irred} show that this reasoning straightforwardly extends to higher orders. By construction, in any irreducible diagram, all ingoing momenta to $O[\dlin_\L]$ as well as $\renormOP{O'}$ are cut at $\L$, leading to the same result for $L_{OO'}$ and $R_{OO'}$. This is completely analogous to the reasoning in \cite{Schmidt:2020viy}.

We conclude that \refeq{toshow} holds for all irreducible contributions.

\subsection{Reducible contributions}
\label{sec:reducible}

We now turn to the \emph{reducible} contributions, which are simply defined as all contributions that are not irreducible. By definition, this means that, in a correlator $\< O^{[n]}(\vk) O'^{[m]}(\vk')\>$, at least one linear propagator is connected only within $O$ or $O'$, so that the entire diagram is $(n+m)/2$-particle-reducible. Note that if $n\neq m$, \emph{all} contributions are reducible.

Let us then consider the reducible contributions to $L_{OO'}$ and $R_{OO'}$. We can consider two types of reducible contributions, based on whether a linear propagator is connected to $O[\dlin_\L]$ or \emph{only to} $\renormOP{O'}$:
\begin{enumerate}
\item If at least one of the outgoing lines of a given propagator is connected to $O[\dlin_\L]$, then, as argued in \cite{Schmidt:2020viy}, this loop involves the cut propagator $\Plin^\L$, and the momentum is restricted to $\leq \L$; for example:
  \be
  \raisebox{-0.0cm}{\includegraphicsbox[scale=1]{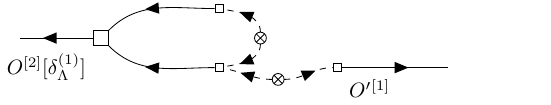}}
  \label{eq:loopT1}
  \ee
\item If both lines of a propagator are connected to $\renormOP{O}$, then the loop momentum is not regularized; for example:
  \be
  \raisebox{-0.0cm}{\includegraphicsbox[scale=1]{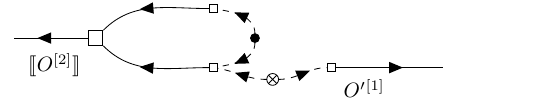}}
  \label{eq:loopT2}
  \ee
  By construction, in the limit of external momentum going to 0, this term is \emph{entirely absorbed} by a counterterm in $\renormOP{O}$. In the case of the diagram above, involving a second-order operator $O^{[2]}$, the counterterm is proportional to $\d$ and, again, completely absorbs this contribution.
\end{enumerate}

To summarize, the second type of reducible contribution vanishes, while the first has a loop momentum that is limited to $\leq \L$, leading to a suppression factor of $\sim \sigma_\L^2 \sim \Lambda^3 \Plin(\L)/2\pi^2$ on large scales. Thus, this correction becomes negligible in the $\L\to 0$ limit considered here. The reducible contributions to $R_{OO'}$ are all of the first type.
We conclude that \emph{all reducible contributions either vanish or are suppressed in the limit indicated in \refeq{toshow}, and thus the asymptotic equality holds also beyond leading order.}

\vspace{.5cm}

To summarize, \refeq{toshow} holds at all orders in perturbation theory and the connection between the two renormalization schemes is realized at $\L \to 0$.

\section{On loop contributions to \texorpdfstring{$\bm{n}$}{n}-point functions in the two approaches}
\label{sec:diffcontrib}

We have argued in the previous section that, in the limit $\Lambda\to 0$ (with $k/\Lambda$ fixed), the bias coefficients $b_O^\L$ run to values that agree with the $n$-point renormalized bias coefficients $b_O^{\rm n-pt}$ used in current EFT analyses.\footnote{Provided that the Gaussian counterterms are included, as in \refeq{gaussiancounter}.} At \emph{finite} $\L$, the $b_O^\L$ assume different values, and hence no longer agree with the $b_O^{\rm n-pt}$. This difference at finite $\Lambda$ is due to the different definitions of operators. In this section we discuss how the UV physics is absorbed by the bias parameters in the two different schemes.  As an example, we consider the three $O=\d^2$ contributions to the 1-loop galaxy power spectrum, which we discuss in turn below. 

\vspace{.5cm}
\textbf{The $\bm{ \boldlangle ( \d)^{(1)}( \d^2)^{(3)} \boldrangle}$ term.}
The first contribution is
\ba
P_{gg}^{\textrm{1-loop, std}}(k) \supset\:& 2 b_\d^{\rm n-pt} b_{\d^2}^{\rm n-pt} \left\< \renormOP{\d}^{(1)}(\vk)\renormOP{\d^2}^{(3)}(\vk')\right\>' \vs
=\:& 4 b_\d^{\rm n-pt} b_{\d^2}^{\rm n-pt}\: \int_p F_2(\vp,\vk-\vp) \Plin(p) \Plin(k) + \mbox{counterterms}.
\label{eq:dd2loop}
\ea
In the standard, $n$-point-renormalized approach, the $k/p\to 0$ limit contribution is completely 
absorbed by the counterterm $(68/21) \sigma^2 \d$ to $\renormOP{\d^2}$; in fact, this is a special case of the diagram in \refeq{loopT2}. The subleading contribution in the limit $k/p\to 0$ is correspondingly absorbed by a counterterm involving $\lapl\d$ (as long as this piece is also properly subtracted). That is, \refeq{dd2loop} disappears completely out of the 1-loop galaxy power spectrum.

On the other hand, when using a finite cutoff, the loop contribution becomes
\ba
P_{gg}^{\textrm{1-loop, finite-scale}}(k) &\supset 2 b_\d^\L b_{\d^2}^\L \left\< \d^{(1)}_\L(\vk) \left(\d^2[\d^{(1)}_\L]\right)^{(3)}(\vk')\right\>' \vs
&= 4 b_\d^\L b_{\d^2}^\L\: \int_p F_2(\vp,\vk-\vp) \Plin^\L(p) \Plin^\L(k).
\label{eq:dd2loopL}
\ea
This contribution [in fact a special case of \refeq{loopT1}] is finite \emph{and is kept as nonzero contribution} to the 1-loop galaxy power spectrum. This explains how, at finite $\L$, a difference between $b_{\d^2}^\L$ and $b_{\d^2}^{\rm n-pt}$ arises. A corresponding finite contribution $\propto (b_\d^\L)^2 b_{\d^2}^\L$ appears in the galaxy bispectrum at 1-loop order,
\be
\raisebox{-0.0cm}{\includegraphicsbox[scale=1]{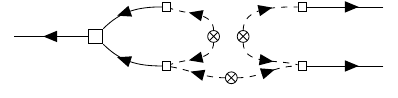}}\,,
\ee
which is controlled by the same $b_{\d^2}^\L$ as \refeq{dd2loopL}.

\vspace{.5cm}
\textbf{The $\bm{ \boldlangle ( \d^2)^{(2)}( \d^2)^{(2)} \boldrangle}$ term.}  
The second contribution we consider is
\ba
&  \raisebox{-0.0cm}{\includegraphicsbox[scale=1]{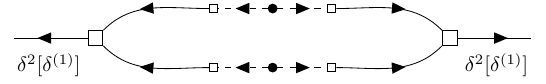}}\vs[5pt]
P_{gg}^{\textrm{1-loop, std}}(k) &\supset (b_{\d^2}^{\rm n-pt})^2 \left\< \renormOP{\d^2}^{(2)}(\vk') \renormOP{\d^2}^{(2)}(\vk')\right\>' \vs
&= 2 (b_{\d^2}^{\rm n-pt})^2\:  \int_{\vp} \Plin(p) \Plin(|\vk-\vp|) + \mbox{counterterms}.
\label{eq:d2d2loop}
\ea
This asymptotes to a UV-sensitive constant in the limit $k/p\to 0$, which is thus to be subtracted, i.e. by removing the shot-noise-like counterterm $\int_{\vp} [\Plin(p)]^2$ from the integral \cite{Assassi2014}. The subleading corrections scale analytically, i.e. as $k^2$, $k^4$, and so on. In order to absorb the UV-dependence of the loop integral at subleading order, these are thus to be correspondingly subtracted and absorbed by higher-derivative stochastic counterterms $P_\eps^{\{2\},\rm n-pt}k^2, P_\eps^{\{4\},\rm n-pt} k^4, ...$ [cf. \refeq{Pstochdef}].
The corresponding contribution from $\< \renormOP{\mathcal{G}_2}^{(2)}(\vk') \renormOP{\d^2}^{(2)}(\vk')\>$ vanishes in the large-scale limit, while the same argument as for $\renormOP{\d^2}$ holds for the subleading contributions.

On the other hand,
\ba
&   \raisebox{-0.0cm}{\includegraphicsbox[scale=1]{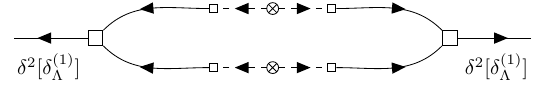}} \vs
P_{gg}^{\textrm{1-loop, finite-scale}}(k) &\supset (b_{\d^2}^{\L})^2 \left\< (\d^2[\d^{(1)}_\L])^{(2)}(\vk') (\d^2[\d^{(1)}_\L])^{(2)}(\vk')\right\>' \vs
&= 2\,(b_{\d^2}^{\L})^2\:  \int_{\vp} \Plin^\L(p) \Plin^\L(|\vk-\vp|)
\label{eq:d2d2loopL}
\ea
is finite and \emph{kept} in the finite-$\L$ renormalization, as in \refeq{dd2loopL}. In an analysis combining the galaxy power spectrum and bispectrum, the constraint on $b_{\d^2}$ from the tree-level bispectrum would then constrain the finite contribution in \refeq{d2d2loopL}. 
The $\L$-dependence of this term is still nonzero however, and needs to be absorbed by the effective shot noise amplitude $P_\eps^\L$.

\vspace{.5cm}
\textbf{The $\bm{ \boldlangle \d^{(2)}( \d^2)^{(2)} \boldrangle}$ term.}  The third and final contribution is
\ba
&   \raisebox{-0.0cm}{\includegraphicsbox[scale=1]{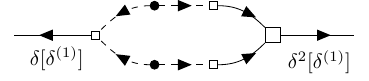}} \vs[5pt]
P_{gg}^{\textrm{1-loop, std}}(k) &\supset 2 b_\d^{\rm n-pt} b_{\d^2}^{\rm n-pt} \left\< \renormOP{\d}^{(2)}(\vk)\renormOP{\d^2}^{(2)}(\vk')\right\>' \vs
&= 4 b_\d^{\rm n-pt} b_{\d^2}^{\rm n-pt}\:  \int_p F_2(\vp,\vk-\vp) \Plin(p) \Plin(|\vk-\vp|) + \mbox{counterterms},
\label{eq:dd2loop2}
\ea
which in the limit $k/p\to 0$ scales as $k^2$, i.e. analytically. Thus, this contribution is UV-sensitive and strictly also to be subtracted by a subleading stochastic counterterm $P_\eps^{(2)} k^2$.

On the other hand,
\ba
&   \raisebox{-0.0cm}{\includegraphicsbox[scale=1]{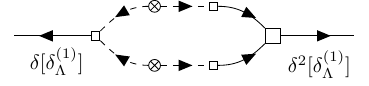}} \vs[5pt]
P_{gg}^{\textrm{1-loop, finite-scale}}(k) &\supset 2 b_\d^{\L} b_{\d^2}^{\L} \left\< \d[\d^{(1)}_\L]^{(2)}(\vk) (\d^2[\d^{(1)}_\L])^{(2)}(\vk')\right\>' \vs
&= 4 b_\d^{\L} b_{\d^2}^{\L}\:  \int_p F_2(\vp,\vk-\vp) \Plin^\L(p) \Plin^\L(|\vk-\vp|) ,
\label{eq:dd2loop2L}
\ea
which is again finite and now requires no further comment.

\vspace{.5cm}

To summarize, the first point to emphasize is that, when implemented consistently, both the usual, $n$-point renormalized bias expansion and the finite-cutoff expansion at $\L$ yield proper UV-insensitive predictions that are independent of non-perturbative small-scale modes. However, in the standard approach, the subtraction of the entire loop terms in \refeq{dd2loop}, \refeq{d2d2loop}, and \refeq{dd2loop2} removes significant contributions from large-scale modes that are in fact under control. This is particularly evident in the subleading contributions in the $k/p\to 0$ limit, for example
\be
k^2 \Plin(k) \int_{\vp} p^{-2} \Plin(p) \quad\mbox{[\refeq{dd2loop}];}\quad
k^2 \int_{\vp} p^{-2} [\Plin(p)]^2 \quad\mbox{[\refeq{d2d2loop}]}.
\label{eq:large_ct}
\ee
As discussed above, these UV-sensitive terms should strictly be removed from the loop integrals, and contribute via the counterterms $b_{\nabla^2\d}^{\rm n-pt}$, $P_\eps^{\{2\},\rm n-pt}$ instead. However, they are large in the sense that they are much larger than $k_{\rm NL}^{-2} \Plin(k)$ and $k_{\rm NL}^{-5}$, which would be the dimensional-analysis estimate of the magnitude of \emph{UV} contributions to each type of counterterms, respectively. Hence, since the corresponding coefficients $b_{\nabla^2\d}^{\rm n-pt}$ and $P_\eps^{\{2\},\rm n-pt}$ now include a significant contribution from low-$k$ modes, one expects that their values are significantly enhanced.
This would also mean that the length scale controlling even higher-order higher-derivative contributions is likewise enhanced.

Working with a finite cutoff, on the other hand, allows for \emph{keeping the large-scale contributions in these loop integrals in a principled way}.
The additional information in these particular loop terms was recently emphasized in Sec.~IV~B of \cite{Cabass:2023nyo}, who considered the covariance of $P_{gg}$ as well.
Fundamentally, the reason that these contributions are absorbed in the standard approach is due to the fact that renormalization is defined via the $k\to 0$ limit of $n$-point functions, which by construction marginalizes over all low-$k$ modes.

At this point, however, we should emphasize that in current EFT analyses the UV-sensitive contributions in \refeq{large_ct} are usually \emph{not} subtracted. This is possible since, while in principle UV-sensitive, they converge to finite values for a $\L$CDM linear power spectrum. The general philosophy appears to have been to remove only \emph{numerically} divergent loop contributions [such as those proportional to $\int_{\vp}^{\infty} \Plin(p)$], while retaining finite contributions such as the subleading contribution to \refeq{d2d2loop} [although the term in \refeq{dd2loop} is usually completely dropped, including numerically finite subleading contributions]. This procedure is fine as long as all counterterm coefficients, i.e. bias and stochastic parameters, are left completely free. As soon as additional information is put on these parameters in the form of priors, great care needs to be taken. E.g., the subleading contribution to \refeq{d2d2loop} is controlled by $(b_{\d^2}^{\rm n-pt})^2$, but its UV-sensitivity has to be absorbed by $P_\eps^{\{2\},\rm n-pt}$. This means that these two parameters are correlated. Priors on the latter parameter that do not consistently take into account this UV sensitivity could bias the inferred value of $b_{\d^2}^{\rm n-pt}$. Similar
arguments apply to higher-order statistics, such as the 1-loop galaxy bispectrum, however involving now a substantially larger bias and stochastic parameter space and even more complex interdependencies.

To reiterate, while priors for a given set of bias and stochastic parameters are usually derived from physical arguments which don't make reference to the details of the UV-sensitive implementation of loop integrals, this is not consistent with the renormalization scheme as currently employed, as these parameters in fact have to absorb UV-sensitive loop contributions, which moreover correlate different parameters. 
  The non-negligible effect of priors in EFTofLSS analyses has indeed been the focus of recent work \cite{Carrilho:2022mon, Simon:2022lde, Donald-McCann:2023kpx}.

  To summarize, the proper $n$-point-function-baed renormalization demands a meticulous (and error-prone) treatment of the basis of operators and loop integrals. For the one-loop bispectrum, for example, it involves subtracting many terms from fourth-order operators. Subleading analytic and higher-derivative contributions (i.e. those scaling as $k^2/p^2$)
  have to be removed from loop integrals.
We emphasize again that all of these subtractions are unnecessary, and all UV sensitivity is absent in the finite-scale renormalization scheme presented in this work. The importance of consistent and physically motivated priors on bias parameters (now at a given scale $\L_*$) applies equally in this case, however.

\section{Discussion and conclusion}
\label{sec:disc}

In this work, we have constructed the renormalization group equations for the galaxy bias and stochastic parameters, Eqs.~(\ref{eq:drun})-(\ref{eq:Grun}). We used the partition function formalism of \cite{Carroll:2013oxa}, generalized to include galaxy bias and stochasticity following \cite{Cabass:2019lqx} to derive the running of those parameters as a function of the smoothing scale $\L$, as modes above $\L$ in the initial conditions are integrated out. We have discussed solutions to this ODE system, indicating that in order to solve the RG evolution at a given order $n$ in the bias expansion, one has to include bias operators up to order $n+2$. We provide solutions for a given set of initial conditions and comment on the stability of the joint first- and second-order bias parameter space under RG flow. 
We also point out
how the running of the bias parameters generates higher-order stochastic contributions. A more detailed investigation of stochasticity is relegated to future work.

We then connect the finite-cutoff bias parameters $b_O^\L$ to their counterparts $b_{O'}^{\rm n-pt}$ defined using the standard $n$-point-function-based renormalization procedure \cite{Assassi2014}. Specifically, we show that the bias parameters of both schemes coincide in the limit $\L \to 0$. The finite cutoff bias parameters, however, allow us to predict additional perturbative contributions to the galaxy $n$-point functions which so far have usually been absorbed, partially or completely, via counterterms (see \refsec{diffcontrib}), in particular into the effective shot noise amplitudes $P_\eps^{\{j\},\rm n-pt}$. This indicates that the finite-cutoff renormalization scheme in principle allows to extract more information from the clustering of biased tracers than has been used in EFT-based analyses so far (see also \cite{Cabass:2023nyo}). This information is available both at the field level and at the level of summary statistics, as we comment on in \refsec{diffcontrib} using the contribution of $b_{\d^2}^\L$ to the power spectrum (at 1-loop level) and the bispectrum (at tree level) as an example.
	A key difference to the standard, $n$-point-function-based renormalization is that the renormalization done here keeps \emph{all linear modes below the cutoff explicit}; subsequently, they can be either marginalized over (for $n$-point functions) or determined by the data (in case of field-level inference). In contrast, the $n$-point-function-based renormalization is done \emph{after marginalizing over all modes, including large scales.}

The difference in the effective shot noise treatment in the standard renormalization approach and the bias expansion at a finite scale (see \refsec{diffcontrib}) can have a particularly large impact for tracers with low ``intrinsic'' shot noise \cite{McQuinn:2018zwa,Obuljen:2022cjo}. Intriguingly, the effective low-$k$ shot noise covariance matrix of dark matter halos appears to have a large eigenvalue associated with $b_{\d^2}$ \cite{Hamaus:2010im}; it is possible that this contribution can be predicted perturbatively when adopting the finite-scale bias expansion.
The actual information gained in the latter over the standard approach will thus depend on the details of the tracer sample, volume, and data vector chosen. This is clearly an issue worthy of further investigation.

Finally, while the finite-scale bias expansion is currently mostly used in field-level forward models \cite{Schmidt:2020viy,Schmidt:2020ovm,Lazeyras:2021dar}, it is worth noting that it can equally be used in standard semi-analytical calculations of $n$-point functions. The presence of the cutoff will however complicate the use of numerical techniques such as \texttt{FFTlog} \cite{Simonovic:2017mhp}. This deserves further investigation, as does the choice of smoother cutoff functions than the Heaviside form adopted here (though a compact support in momentum space is essential to ensure UV insensitivity).

\acknowledgments
HR is supported by the Deutsche Forschungsgemeinschaft under Germany's Excellence Strategy EXC 2094 `ORIGINS'. (No.\,390783311). 
We thank Ivana Babi\'{c}, Giovanni Cabass, Stephen Chen, Mathias Garny, Julia Stadler, Beatriz Tucci and Rodrigo Voivodic for discussions,
and Ivana Babi\'{c}, Julia Stadler and Mark Wise for feedback on the paper. 

\appendix

\section{Shell integrals}  \label{app:shell}
In this appendix, we provide a complete calculation of the shell operators and integrals from \refsec{running}.
\subsection{Shell operators} \label{app:shellOP}

We display the shell expansion of the operators defined in Eq.~(\ref{eq:Oshell}) up to third order in perturbation theory. Notice that only the contributions that have exactly two powers $\dlinshell$ will contribute to $\Shell^2$ and therefore to the bias running. However, for completeness, we display 
the full expressions for the shell expansion.
\begin{itemize}

    \item $O = \d$:
{\small \bea
&&\d[\dlin_\L+\dlinshell](\vk) =  \d[\dlin_\L](\vk) + \dlinshell(\vk) \\
&&\quad + \int_{\vp_1,\vp_2} \dirac(\vk-\vp_{12}) F_2(\vp_1,\vp_2) \left[ 2\dlinshell(\vp_1)\dlin_\L(\vp_2) + \dlinshell(\vp_1)\dlinshell(\vp_2) \right] \vs
&&\quad + \int_{\vp_1,\vp_2,\vp_3} \dirac(\vk-\vp_{123})\, F_3(\vp_1,\vp_2,\vp_3) \vs 
&&\quad  \times \left[ 3\,\dlinshell(\vp_1)\dlin_\L(\vp_2)\dlin_\L(\vp_3) + 3\,\dlinshell(\vp_1)\dlinshell(\vp_2)\dlin_\L(\vp_3) + \dlinshell(\vp_1)\dlinshell(\vp_2)\dlinshell(\vp_3)\right] \vs
&&\quad + \int_{\vp_1,\vp_2,\vp_3,\vp_4} \dirac(\vk-\vp_{1234})\, F_4(\vp_1,\vp_2,\vp_3,\vp_4) \left[ 4\,\dlinshell(\vp_1)\dlin_\L(\vp_2)\dlin_\L(\vp_3)\dlin_\L(\vp_4)   \right. \vs
&&\quad +   6\,\dlinshell(\vp_1)\dlinshell(\vp_2)\dlin_\L(\vp_3)\dlin_\L(\vp_4) + 4\,\dlinshell(\vp_1)\dlinshell(\vp_2)\dlinshell(\vp_3)\dlin_\L(\vp_4) \vs
&&\quad + \left. \dlinshell(\vp_1)\dlinshell(\vp_2)\dlinshell(\vp_3)\dlinshell(\vp_4)\right]\,. \nonumber
\eea}

\item $O = \d^2$:
{\small \bea
&&\d^2[\dlin_\L+\dlinshell](\vk) =  \d^2[\dlin_\L](\vk) + \int_{\vp_1,\vp_2} \dirac(\vk-\vp_{12}) \left[ 2\dlinshell(\vp_1)\dlin_\L(\vp_2) + \dlinshell(\vp_1)\dlinshell(\vp_2) \right]\vs 
&&\quad + 2\int_{\vp_1,\vp_2,\vp_3} \dirac(\vk-\vp_{123}) F_2(\vp_1,\vp_2)  \left[ 2\dlinshell(\vp_1)\dlin_\L(\vp_2)\dlin_\L(\vp_3) + \dlin_\L(\vp_1)\dlin_\L(\vp_2)\dlinshell(\vp_3) \right. \vs
&&\quad + \left. 2\dlin_\L(\vp_1)\dlinshell(\vp_2)\dlinshell(\vp_3) + \dlinshell(\vp_1)\dlinshell(\vp_2)\dlin_\L(\vp_3) +  \dlinshell(\vp_1)\dlinshell(\vp_2)\dlinshell(\vp_3)  \right] \vs 
&&\quad + \int_{\vp_1,\vp_2,\vp_3,\vp_4} \dirac(\vk-\vp_{1234}) F_2(\vp_1,\vp_2)F_2(\vp_3,\vp_4)   \left[ 4\, \dlinshell(\vp_1)\dlin_\L(\vp_2)\dlin_\L(\vp_3)\dlin_\L(\vp_4) \right.  \vs
&&\quad + 4\, \dlinshell(\vp_1)\dlin_\L(\vp_2)\dlinshell(\vp_3)\dlin_\L(\vp_4) + 2\, \dlinshell(\vp_1)\dlinshell(\vp_2)\dlin_\L(\vp_3)\dlin_\L(\vp_4)  \vs
&& \quad \left. + 4\, \dlinshell(\vp_1)\dlinshell(\vp_2)\dlinshell(\vp_3)\dlin_\L(\vp_4) + \, \dlinshell(\vp_1)\dlinshell(\vp_2)\dlinshell(\vp_3)\dlinshell(\vp_4)\right] \vs 
&&\quad + 2\int_{\vp_1,\vp_2,\vp_3,\vp_4} \dirac(\vk-\vp_{1234}) F_3(\vp_1,\vp_2,\vp_3)  \left[ 3\dlinshell(\vp_1)\dlin_\L(\vp_2)\dlin_\L(\vp_3)\dlin_\L(\vp_4)\right. \vs
&&\quad + \dlin_\L(\vp_1)\dlin_\L(\vp_2)\dlin_\L(\vp_3)\dlinshell(\vp_4) +  3\dlinshell(\vp_1)\dlinshell(\vp_2)\dlin_\L(\vp_3)\dlin_\L(\vp_4)  \vs
&&\quad  + 3 \dlinshell(\vp_1)\dlin_\L(\vp_2)\dlin_\L(\vp_3)\dlinshell(\vp_4) +  \dlinshell(\vp_1)\dlinshell(\vp_2)\dlinshell(\vp_3)\dlin_\L(\vp_4)  \vs
&& \quad  +  3\dlin_\L(\vp_1)\dlinshell(\vp_2)\dlinshell(\vp_3)\dlinshell(\vp_4) +  \left. \dlinshell(\vp_1)\dlinshell(\vp_2)\dlinshell(\vp_3)\dlinshell(\vp_4)  \right]
\eea}

\item $O = \d^3$: 
{\small \bea
&&\d^3[\dlin_\L+\dlinshell](\vk) =  \d^3[\dlin_\L](\vk) + \int_{\vp_1,\vp_2,\vp_3} \dirac(\vk-\vp_{123}) \vs 
&&\quad \times \left[ 3\dlinshell(\vp_1)\dlin_\L(\vp_2)\dlin_\L(\vp_3) + 3\dlinshell(\vp_1)\dlinshell(\vp_2)\dlin_\L(\vp_3) + \dlinshell(\vp_1)\dlinshell(\vp_2)\dlinshell(\vp_3) \right]  \vs 
&& \quad + 3 \int_{\vp_1,\vp_2,\vp_3,\vp_4} \dirac(\vk-\vp_{1234}) F_2(\vp_1,\vp_2) \left[ 2\,\dlinshell(\vp_1)\dlin_\L(\vp_2)\dlin_\L(\vp_3)\dlin_\L(\vp_4)\right. \vs
&& \quad + 2\dlin_\L(\vp_1)\dlin_\L(\vp_2)\dlinshell(\vp_3)\dlin_\L(\vp_4) + \dlinshell(\vp_1)\dlinshell(\vp_2)\dlin_\L(\vp_3)\dlin_\L(\vp_4) \vs 
&& \quad   + \dlin_\L(\vp_1)\dlin_\L(\vp_2)\dlinshell(\vp_3)\dlinshell(\vp_4) + 4\dlin_\L(\vp_1)\dlinshell(\vp_2)\dlin_\L(\vp_3)\dlinshell(\vp_4) \\
&& \quad   + 2\,\dlin_\L(\vp_1)\dlinshell(\vp_2)\dlinshell(\vp_3)\dlinshell(\vp_4) + 2\,\dlinshell(\vp_1)\dlinshell(\vp_2)\dlin_\L(\vp_3)\dlinshell(\vp_4)  \vs
&& \quad   \left.+ \dlinshell(\vp_1)\dlinshell(\vp_2)\dlinshell(\vp_3)\dlinshell(\vp_4)\right] \nonumber
\eea}

\item $O = \d^4$:
{\small \bea
&&\d^4[\dlin_\L+\dlinshell](\vk) =  \d^4[\dlin_\L](\vk) + \int_{\vp_1,\vp_2,\vp_3,\vp_4} \dirac(\vk-\vp_{1234})\left[ 4\, \dlinshell(\vp_1)\dlin_\L(\vp_2)\dlin_\L(\vp_3)\dlin_\L(\vp_4)  \right. \vs
&& \quad  +   6\,\dlinshell(\vp_1)\dlinshell(\vp_2)\dlin_\L(\vp_3)\dlin_\L(\vp_4) + 4 \,\dlinshell(\vp_1)\dlinshell(\vp_2)\dlinshell(\vp_3)\dlin_\L(\vp_4) \\
&& \quad  \left. + \dlinshell(\vp_1)\dlinshell(\vp_2)\dlinshell(\vp_3)\dlinshell(\vp_4) \right] \,,\nonumber
\eea}

\item $O = \G_2$:

{\small \bea
&&\G_2[\dlin_\L+\dlinshell](\vk) =  \G_2[\dlin_\L](\vk) \vs
&& \quad + \int_{\vp_1,\vp_2} \dirac(\vk-\vp_{12})  \s^2_{\vp_1,\vp_2} \left[ 2\dlinshell(\vp_1)\dlin_\L(\vp_2) + \dlinshell(\vp_1)\dlinshell(\vp_2) \right]\vs 
&&\quad + 2\int_{\vp_1,\vp_2,\vp_3} \dirac(\vk-\vp_{123}) F_2(\vp_1,\vp_2)\s^2_{\vp_1+\vp_2,\vp_3}  \vs
&&\quad  \times \left[ 2\dlinshell(\vp_1)\dlin_\L(\vp_2)\dlin_\L(\vp_3) + \dlin_\L(\vp_1)\dlin_\L(\vp_2)\dlinshell(\vp_3) \right. \vs
&&\quad + \left. 2\dlin_\L(\vp_1)\dlinshell(\vp_2)\dlinshell(\vp_3) + \dlinshell(\vp_1)\dlinshell(\vp_2)\dlin_\L(\vp_3) +  \dlinshell(\vp_1)\dlinshell(\vp_2)\dlinshell(\vp_3)  \right] \vs 
&&\quad + \int_{\vp_1,\vp_2,\vp_3,\vp_4} \dirac(\vk-\vp_{1234}) F_2(\vp_1,\vp_2)F_2(\vp_3,\vp_4)   \s^2_{\vp_1+\vp_2,\vp_3+\vp_4} \vs
&& \quad \left[ 4\, \dlinshell(\vp_1)\dlin_\L(\vp_2)\dlin_\L(\vp_3)\dlin_\L(\vp_4)\right. \vs 
&&\quad + 4\, \dlinshell(\vp_1)\dlin_\L(\vp_2)\dlinshell(\vp_3)\dlin_\L(\vp_4)  + 2\, \dlinshell(\vp_1)\dlinshell(\vp_2)\dlin_\L(\vp_3)\dlin_\L(\vp_4)  \vs
&&\quad  \left.+ 4\, \dlinshell(\vp_1)\dlinshell(\vp_2)\dlinshell(\vp_3)\dlin_\L(\vp_4) + \, \dlinshell(\vp_1)\dlinshell(\vp_2)\dlinshell(\vp_3)\dlinshell(\vp_4)\right] \vs 
&&\quad + 2\int_{\vp_1,\vp_2,\vp_3,\vp_4} \dirac(\vk-\vp_{1234}) F_3(\vp_1,\vp_2,\vp_3) \s^2_{\vp_1+\vp_2+\vp_3,\vp_4} \left[ 3\dlinshell(\vp_1)\dlin_\L(\vp_2)\dlin_\L(\vp_3)\dlin_\L(\vp_4)  \right. \vs
&&\quad  + \dlin_\L(\vp_1)\dlin_\L(\vp_2)\dlin_\L(\vp_3)\dlinshell(\vp_4) +  3\dlinshell(\vp_1)\dlinshell(\vp_2)\dlin_\L(\vp_3)\dlin_\L(\vp_4)   \vs
&&\quad  + 3 \dlinshell(\vp_1)\dlin_\L(\vp_2)\dlin_\L(\vp_3)\dlinshell(\vp_4) +  \dlinshell(\vp_1)\dlinshell(\vp_2)\dlinshell(\vp_3)\dlin_\L(\vp_4) \vs
&&\quad  \left. +    \dlinshell(\vp_1)\dlinshell(\vp_2)\dlinshell(\vp_3)\dlinshell(\vp_4)  \right]
\eea}

\item $O = \G_2\d$:

{\small \bea
&&\G_2\d[\dlin_\L+\dlinshell](\vk) =  \G_2\d[\dlin_\L](\vk) \vs
&&\quad + \int_{\vp_1,\vp_2,\vp_3} \dirac(\vk-\vp_{123})  \s^2_{\vp_1,\vp_2}   \left[  2\dlinshell(\vp_1)\dlin_\L(\vp_2)\dlin_\L(\vp_3) + \dlin_\L(\vp_1)\dlin_\L(\vp_2)\dlinshell(\vp_3)  \right. \vs
&&\quad  \left. + \dlinshell(\vp_1)\dlinshell(\vp_2)\dlin_\L(\vp_3) + 2\dlinshell(\vp_1)\dlin_\L(\vp_2)\dlinshell(\vp_3) + \dlinshell(\vp_1)\dlinshell(\vp_2)\dlinshell(\vp_3) \right]  \vs 
&&\quad +  \int_{\vp_1,\vp_2,\vp_3,\vp_4} \dirac(\vk-\vp_{1234}) \s^2_{\vp_1,\vp_2}F_2(\vp_3,\vp_4)   \left[ 2\,\dlinshell(\vp_1)\dlin_\L(\vp_2)\dlin_\L(\vp_3)\dlin_\L(\vp_4)\right. \vs
&&\quad + 2\dlin_\L(\vp_1)\dlin_\L(\vp_2)\dlinshell(\vp_3)\dlin_\L(\vp_4) + \dlinshell(\vp_1)\dlinshell(\vp_2)\dlin_\L(\vp_3)\dlin_\L(\vp_4)  \vs 
&&\quad   + \dlin_\L(\vp_1)\dlin_\L(\vp_2)\dlinshell(\vp_3)\dlinshell(\vp_4) + 4\dlin_\L(\vp_1)\dlinshell(\vp_2)\dlin_\L(\vp_3)\dlinshell(\vp_4) \vs
&&\quad  + 2\,\dlin_\L(\vp_1)\dlinshell(\vp_2)\dlinshell(\vp_3)\dlinshell(\vp_4) + 2\,\dlinshell(\vp_1)\dlinshell(\vp_2)\dlin_\L(\vp_3)\dlinshell(\vp_4)  \vs
&&\quad \left.  + \dlinshell(\vp_1)\dlinshell(\vp_2)\dlinshell(\vp_3)\dlinshell(\vp_4)\right] \vs 
&&\quad + 2 \int_{\vp_1,\vp_2,\vp_3,\vp_4} \dirac(\vk-\vp_{1234}) F_2(\vp_1,\vp_2) \s^2_{\vp_{12},\vp_3}   \left[ 2\,\dlinshell(\vp_1)\dlin_\L(\vp_2)\dlin_\L(\vp_3)\dlin_\L(\vp_4) \right.\vs
&&\quad +2\dlin_\L(\vp_1)\dlin_\L(\vp_2)\dlinshell(\vp_3)\dlin_\L(\vp_4) + \dlin_\L(\vp_1)\dlin_\L(\vp_2)\dlin_\L(\vp_3)\dlinshell(\vp_4)  \vs 
&&\quad   + \dlinshell(\vp_1)\dlinshell(\vp_2)\dlin_\L(\vp_3)\dlin_\L(\vp_4) + 2\dlinshell(\vp_1)\dlin_\L(\vp_2)\dlinshell(\vp_3)\dlin_\L(\vp_4) \vs
&&\quad  + 2\dlinshell(\vp_1)\dlin_\L(\vp_2)\dlin_\L(\vp_3)\dlinshell(\vp_4) + \dlin_\L(\vp_1)\dlin_\L(\vp_2)\dlinshell(\vp_3)\dlinshell(\vp_4)  \vs
&&\quad + 2\,\dlin_\L(\vp_1)\dlinshell(\vp_2)\dlinshell(\vp_3)\dlinshell(\vp_4)+ \,\dlinshell(\vp_1)\dlinshell(\vp_2)\dlin_\L(\vp_3)\dlinshell(\vp_4) \\
&&\quad  \left.   + \,\dlinshell(\vp_1)\dlinshell(\vp_2)\dlinshell(\vp_3)\dlin_\L(\vp_4)  + \dlinshell(\vp_1)\dlinshell(\vp_2)\dlinshell(\vp_3)\dlinshell(\vp_4)\right]  \nonumber
\eea}

\item $O = \G_3$. We do not display the shell opetaror for the third order Galileon \refeq{galileon3}, which is rather long. The Galileon operators are non-renormalizable \cite{Nicolis:2008in} and therefore do not act as sources for other operators \cite{Assassi2014}.

\item $O = \Gamma_3$:
{\small \bea
&&\Gamma_3[\dlin_\L+\dlinshell](\vk) =  \Gamma_3[\dlin_\L](\vk) \vs 
&&\quad  + 2\int_{\vp_1,\vp_2,\vp_3} \dirac(\vk-\vp_{123}) \left[F_2(\vp_1,\vp_2)-G_2(\vp_1,\vp_2)\right]\s^2_{\vp_1+\vp_2,\vp_3} \vs
&&\quad   \times \left[ 2\dlinshell(\vp_1)\dlin_\L(\vp_2)\dlin_\L(\vp_3) + \dlin_\L(\vp_1)\dlin_\L(\vp_2)\dlinshell(\vp_3) \right. \vs
&&\quad  + \left. 2\dlin_\L(\vp_1)\dlinshell(\vp_2)\dlinshell(\vp_3) + \dlinshell(\vp_1)\dlinshell(\vp_2)\dlin_\L(\vp_3) +  \dlinshell(\vp_1)\dlinshell(\vp_2)\dlinshell(\vp_3)  \right] \vs 
&&\quad  + \int_{\vp_1,\vp_2,\vp_3,\vp_4} \dirac(\vk-\vp_{1234}) \left[F_2(\vp_1,\vp_2)F_2(\vp_3,\vp_4) - G_2(\vp_1,\vp_2)G_2(\vp_3,\vp_4)\right]   \s^2_{\vp_1+\vp_2,\vp_3+\vp_4} \vs 
&&\quad   \times \left[ 4\, \dlinshell(\vp_1)\dlin_\L(\vp_2)\dlin_\L(\vp_3)\dlin_\L(\vp_4)  + 4\, \dlinshell(\vp_1)\dlin_\L(\vp_2)\dlinshell(\vp_3)\dlin_\L(\vp_4)  \right. \vs
&&\quad   + 2\, \dlinshell(\vp_1)\dlinshell(\vp_2)\dlin_\L(\vp_3)\dlin_\L(\vp_4)+ 4\, \dlinshell(\vp_1)\dlinshell(\vp_2)\dlinshell(\vp_3)\dlin_\L(\vp_4) \vs
&&\quad   \left. + \, \dlinshell(\vp_1)\dlinshell(\vp_2)\dlinshell(\vp_3)\dlinshell(\vp_4)\right] \vs 
&&\quad  + 2\int_{\vp_1,\vp_2,\vp_3,\vp_4} \dirac(\vk-\vp_{1234}) \left[F_3(\vp_1,\vp_2,\vp_3) - G_3(\vp_1,\vp_2,\vp_3)\right] \s^2_{\vp_1+\vp_2+\vp_3,\vp_4} \vs 
&&\quad   \times \left[ 3\dlinshell(\vp_1)\dlin_\L(\vp_2)\dlin_\L(\vp_3)\dlin_\L(\vp_4) + \dlin_\L(\vp_1)\dlin_\L(\vp_2)\dlin_\L(\vp_3)\dlinshell(\vp_4) \right.  \vs
&&\quad   +  3\dlinshell(\vp_1)\dlinshell(\vp_2)\dlin_\L(\vp_3)\dlin_\L(\vp_4) + 3 \dlinshell(\vp_1)\dlin_\L(\vp_2)\dlin_\L(\vp_3)\dlinshell(\vp_4) \\
&&\quad    \left. +  \dlinshell(\vp_1)\dlinshell(\vp_2)\dlinshell(\vp_3)\dlin_\L(\vp_4) +  \dlinshell(\vp_1)\dlinshell(\vp_2)\dlinshell(\vp_3)\dlinshell(\vp_4)  \right] \nonumber
\eea}
\end{itemize}

\subsection{The $\Shell^2_O$ integrals}  \label{app:shellInt}

We now display the shell integrals $\Shell^2_O$ defined in Eqs.~(\ref{eq:shell}) and (\ref{eq:shell2}),
which correspond to taking the expectation value over $\dlinshell$ of the contributions $\O([\dlinshell]^2)$ to the expansion defined in \refapp{shellOP}.
As we discuss in Sec.~\ref{sec:running}, these expectation values will contribute to the bias RG flow.
Term by term, we find: 
\begin{itemize}
    \item $O = \d$:  
{\small	\bea
	\Shell^2_{\d} [\dlin_\L](\vk) 
	&=&  3\dlin_\L(\vk) \int_{\vp} F_3(\vp,-\vp,\vk)  P_{\rm shell}(p) 
	\vs
	&& \quad + \,6 \int_{\vp,\vp_1,\vp_2} \dirac(\vk-\vp_{12}) F_4(\vp,-\vp,\vp_1,\vp_{2})  P_{\rm shell}(p) \dlin_\L(\vp_1)\dlin_\L(\vp_{2}) \vs
	&& \quad +\, \O\left[\left(\dlin_\L\right)^3\right] \vs 
	&\stackrel{k\ll p}{=}&  -\frac{61}{315} k^2 \dlin_\L(\vk) \int \frac{p^2 dp}{2\pi^2} \,   \frac{P_{\rm shell}(p)}{p^2} +\, \O\left[\nabla^2 \left(\dlin_\L\right)^2\right]  \,,
	\eea}where $\O[\nabla^2(\dlin_\L)^2]$ represents higher-derivative contributions that are also higher order in $\dlin_\L$. Notice that a higher-order expansion of $\d$ will likewise only lead to contributions to higher-derivative operators, since the momentum structure of the kernels is always of the form $F_{\ell}(p,-p,\dots)$.

    \item $O = \d^2$:
{\small\bea 
  \Shell^2_{\d^2} [\dlin_\L](\vk) &=&   4\dlin_\L(\vk) \int_{\vp} F_2(\vk,\vp)  P_{\rm shell}(p) \\
  &+& \quad  6 \int_{\vp,\vp_1,\vp_2} \dirac(\vk-\vp_{12}) \left[ F_3(\vp,-\vp, \vp_1) +  F_3(\vp,\vp_1, \vp_2)\right] P_{\rm shell}(p) \dlin_\L(\vp_1)\dlin_\L(\vp_{2}) \vs
  &+&  \quad 4 \int_{\vp,\vp_1,\vp_2} \dirac(\vk-\vp_{12}) F_2(\vp,\vp_1)F_2(-\vp,\vp_2)  P_{\rm shell}(p) \dlin_\L(\vp_1)\dlin_\L(\vp_{2}) \vs
  &+& \quad  \O\left[\left(\dlin_\L\right)^3\right] \,. \nonumber
\eea}where we can immediately see that the first term will contribute to $\d$
\bea
  4\dlin_\L(\vk) \int_{\vp} F_2(\vk,\vp)  P_{\rm shell}(p)  &=& \frac{68}{21} \dlin_\L(\vk) \int \frac{p^2 dp}{2\pi^2} \, P_{\rm shell}(p)   \,.
\eea
One of the $F_3$ terms is also of the form $F_n(\vp,-\vp,\dots)$ such that
{\small \bea
   \int_{\vp,\vp_1,\vp_2}  \dirac(\vk-\vp_{12}) F_3(\vp,-\vp, \vp_1)  P_{\rm shell}(p) \dlin_\L(\vp_1)\dlin_\L(\vp_{2}) \propto&& \\ \int_{\vp,\vp_1,\vp_2} \dirac(\vk-\vp_{12}) && \left(\frac{p_1}{p}\right)^2 P_{\rm shell}(p) \dlin_\L(\vp_1)\dlin_\L(\vp_{2}) \nonumber
\eea}and it will renormalize $\d\nabla^2\d$. For the $22$ term, we find
\bea \label{eq:F2F2} 
  &&4 \int_{\vp}  F_2(\vp,\vp_1)F_2(-\vp,\vp_2)  P_{\rm shell}(p)  = \int \frac{p^2 dp}{2 \pi^2 }P_{\rm shell}(p) \times \\
 &&\quad \left[ \frac{1916}{735} + \frac{32}{735}\frac{(\vp_1\cdot \vp_2)^2}{p_1^2 p_2^2} - \frac{1}{3}\frac{(\vp_1\cdot \vp_2)}{  p^2} - \frac{1}{3}\frac{(\vp_1\cdot \vp_2)}{  p_1^2} - \frac{1}{3}\frac{(\vp_1\cdot \vp_2)}{  p_2^2}  - \frac{1}{3}\frac{p^2(\vp_1\cdot \vp_2)}{ p_1^2 p_2^2}  \right]\,.\nonumber
\eea
Notice that the first term will source contributions to $\d^2$ and the second to $\G_2$ and $\d^2$. The third term in Eq.~(\ref{eq:F2F2}) sources the higher-derivative term $\nabla_i \d \, \nabla_j \d$. However, the last three terms in Eq.~(\ref{eq:F2F2}) do {\it not} respect Galilean invariance, since they come from displacements terms like $\partial_i \Phi \partial_j \Phi$ and $\partial_i \Phi \partial_j \d$. We comment more on them later.
We now turn to $F_3(\vp,\vp_1, \vp_2)$, which we expand using $p \to \infty$ to get 
\bea\label{eq:F3} 
&&6 \int_{\vp,\vp_1}  F_3(\vp,\vp_1, \vp_2)  P_{\rm shell}(p) \stackrel{p \to \infty}{\approx}  \int \frac{p^2 dp}{2 \pi^2 }P_{\rm shell}(p) \times \\
&&\quad \left[ \frac{344}{105} + \frac{314}{315}\frac{(\vp_1\cdot \vp_2)^2}{p_1^2 p_2^2} + \frac{41}{21}\frac{(\vp_1\cdot \vp_2)}{  p_1^2}  + \frac{41}{21}\frac{(\vp_1\cdot \vp_2)}{  p_2^2} + \frac{1}{3}\frac{p^2(\vp_1\cdot \vp_2)}{ p_1^2 p_2^2} + \O\left(\frac{p_1^2}{p^{2}},\frac{p_2^2}{p^{2}}\right)  \right]\,.\nonumber
\eea
We can see that the term proportional to $(\vp_1\cdot \vp_2)/(p_1^2 p_2^2)$ will will cancel out the term from Eq.~(\ref{eq:F2F2}) and therefore the non-Galilean-invariant term $\partial_i \Phi \partial_j \Phi$ disappears. 
We can then write
{\small\bea
  \int_{\vp,\vp_1,\vp_2}&& \dirac(\vk-\vp_{12}) \left[ 4 F_2(\vp,\vp_1)F_2(-\vp,\vp_2) +  6 F_3(\vp,\vp_1, \vp_2)  \right]  P_{\rm shell}(p) \dlin_\L(\vp_1)\dlin_\L(\vp_{2})   \vs
   \quad &=&\int \frac{p^2 dp}{2 \pi^2 }P_{\rm shell}(p) \int_{\vp_1,\vp_2}\dirac(\vk-\vp_{12})\dlin_\L(\vp_1)\dlin_\L(\vp_{2})   \vs
   && \quad \times\left[ \frac{4324}{735} + \frac{2294}{2205}\frac{(\vp_1\cdot \vp_2)^2}{p_1^2 p_2^2}  + \frac{34}{21}\frac{(\vp_1\cdot \vp_2)}{  p_1^2} + \frac{34}{21}\frac{(\vp_1\cdot \vp_2)}{  p_2^2}    \right]\vs
   \quad &=& \left[ \frac{68}{21}\d^{(2)}(\vk) + \frac{8126}{2205}(\d^2)^{(2)}(\vk) +\frac{254}{2205}\G_2^{(2)}(\vk) \right] \int \frac{p^2 dp}{2 \pi^2 }P_{\rm shell}(p)\,,
\eea}where we find that it will contribute to $\d,\,\d^2,\,\G_2$. Putting all terms together:
\bea \label{eq:Sd2A}
 \Shell^2_{\d^2}[\dlin_\L] (\vk) &=& \left[ \frac{68}{21}\d^{(1+2)}(\vk) + \frac{8126}{2205}\left[\d^{2}(\vk)\right]^{(2)} +\frac{254}{2205}\G_2^{(2)}(\vk) \right] \int \frac{p^2 dp}{2 \pi^2 }P_{\rm shell}(p) \\ 
 && \quad \quad \quad \quad \quad \quad \quad \quad \quad \quad \quad \quad \quad  + \,\mbox{higher derivative (h.d.)} + \O\left[\left(\dlin_\L\right)^3\right]\,. \nonumber
\eea
An important result is that the coefficients of $\d^{(1)}(\vk)$ and $\d^{(2)}(\vk)$ are the same, which can also be understood as a consequence of Galilean invariance. 

    \item $O = \d^3$:
{\small\bea 
\Shell^2_{\d^3} [\dlin_\L](\vk) &=&  3\dlin_\L(\vk) \int_{\vp} P_{\rm shell}(p)+ 3 \int_{\vp,\vp_1,\vp_2}\dirac(\vk-\vp_{12})  F_2(\vp_1,\vp_2)  P_{\rm shell}(p) \dlin_\L(\vp_1)\dlin_\L( \vp_{2}) \vs
&& + 3 \int_{\vp,\vp_1,\vp_2} \dirac(\vk-\vp_{12}) F_2(\vp_1,\vp)  P_{\rm shell}(p) \dlin_\L(\vp_1)\dlin_\L(\vp_{2}) +  \O\left[\left(\dlin_\L\right)^3\right]\\
&=&  3 \left(\d[\d_\L^{(1)}]\right)^{(1+2)}(\vk)  \int_{\vp} P_{\rm shell}(p) + \frac{17}{7}  \left(\d^2[\d_\L^{(1)}]\right)^{(2)}(\vk)   \int_{\vp}   P_{\rm shell}(p) +  \O\left[\left(\dlin_\L\right)^3\right] \,, \nonumber
\eea}where we see that the first two terms contribute to $\d[\d_\L^{(1)}]$ (at leading and second order) and the final term contributes to $\d^2[\d_\L^{(1)}]$. We highlight again that the contributions to  $\left(\d[\d_\L^{(1)}]\right)^{(1)}$ and $\left(\d[\d_\L^{(1)}]\right)^{(2)}$ have the same coefficient. 

    \item $O = \d^4$. In order to understand the expected behaviour of fourth-order terms, we calculate the shell contribution from $\d^4$:
\bea
\Shell^2_{\d^4} [\dlin_\L](\vk) &=&  6\int_{\vp,\vp_1} P_{\rm shell}(p) \dlin_\L(\vp_1)\dlin_\L(\vk-\vp_1) + \O\left[\left(\dlin_\L\right)^3\right] \vs 
&=& 6\left(\d^2[\d_\L^{(1)}]\right)^{(2)}(\vk)\int_{\vp} P_{\rm shell}(p)  + \O\left[\left(\dlin_\L\right)^3\right] \,,
\eea
which contributes only to $\d^2[\d_\L^{(1)}]$. We notice that the operators of order $n$ can only contribute to operators starting from order $n-2$.

    \item $O = \G_2$:  
\bea
 \Shell^2_{\G_2}[\dlin_\L] (\vk) &=&  4\dlin_\L(\vk) \int_{\vp}  \s^2_{\vk+\vp,-\vp}F_2(\vk,\vp)  P_{\rm shell}(p) \vs 
 &+& 4 \int_{\vp,\vp_1,\vp_2} \dirac(\vk-\vp_{12}) F_2(\vp,\vp_1)F_2(-\vp,\vp_2) \vs
 && \hspace{5cm}\times \s^2_{\vp_2-\vp,\vp+\vp_1}P_{\rm shell}(p) \dlin_\L(\vp_1)\dlin_\L(\vp_{2}) \vs
  &+& 6 \int_{\vp,\vp_1,\vp_2} \dirac(\vk-\vp_{12}) \s^2_{\vp_1,\vp_2} F_3(\vp,-\vp, \vp_1)  P_{\rm shell}(p) \dlin_\L(\vp_1)\dlin_\L(\vp_{2}) \vs 
  &+&  6 \int_{\vp,\vp_1,\vp_1} \dirac(\vk-\vp_{12}) \s^2_{-\vp,\vk+\vp}  F_3(\vp,\vp_1,\vp_2)  P_{\rm shell}(p) \dlin_\L(\vp_1)\dlin_\L(\vp_{2}) \vs
 &\stackrel{k\ll p}{=}&   \dlin_\L(\vk) \int_{\vp}\,   P_{\rm shell}(p)\left[ -\frac{32}{21}\frac{k^2}{p^2} + \dots \right]  + k^2[\dots] \,.
\eea 
Notice that all terms have a structure like $\s^2_{\vp + \dots,-\vp+\dots}$ or $F_3(\vp,-\vp, \vp_1)$. Both types of terms scale as $p^2$ and only contribute to higher-derivative terms. That is a consequence of the non-renormalization theorem for Galileon operators \cite{Nicolis:2008in,Assassi2014}. 

    \item $O = \G_2\d$:
\bea
  \Shell^2_{\G_2\d} [\dlin_\L](\vk) &=&   2\dlin_\L(\vk) \int_{\vp} \s^2_{\vp,\vk}  P_{\rm shell}(p) \vs
  &+& 4 \int_{\vp,\vp_1,\vp_2} \dirac(\vk-\vp_{12}) \s^2_{\vp,\vp_1}F_2(\vp_2,-\vp)  P_{\rm shell}(p) \dlin_\L(\vp_1)\dlin_\L(\vk-\vp_1)    \vs 
  &+& 4 \int_{\vp,\vp_1,\vp_2} \dirac(\vk-\vp_{12}) \s^2_{\vp + \vp_{1},\vp}F_2(\vp,\vp_1)  P_{\rm shell}(p) \dlin_\L(\vp_1)\dlin_\L(\vp_2)  \vs
  &+& 4 \int_{\vp,\vp_1,\vp_2} \dirac(\vk-\vp_{12}) \s^2_{\vp + \vp_{1},\vp_2}F_2(\vp,\vp_1)  P_{\rm shell}(p) \dlin_\L(\vp_1)\dlin_\L(\vp_2) \vs
  &+& 2 \int_{\vp,\vp_1,\vp_2} \dirac(\vk-\vp_{12}) \s^2_{\vk,\vp}F_2(\vp_1,\vp_2)  P_{\rm shell}(p) \dlin_\L(\vp_1)\dlin_\L(\vp_2)\,. \label{eq:shellGd}
\eea
The first term
\bea
2\dlin_\L(\vk) \int_{\vp} \s^2_{\vk,\vp} P_{\rm shell}(p) =-\frac{4}{3}\dlin_\L(\vk) \int_{\vp} P_{\rm shell}(p)\,,
\eea
and the last term
\bea
2\int_{\vp} \s^2_{\vk,\vp} P_{\rm shell}(p) \int_{\vp_1,\vp_2} \dirac(\vk-\vp_{12}) F_2(\vp_1,\vp_2)   \dlin_\L(\vp_1)\dlin_\L(\vp_2) &\\
= -\frac{4}{3}\int_{\vp} P_{\rm shell}(p) \int_{\vp_1,\vp_2} \dirac(\vk-\vp_{12})  F_2(\vp_1,\vp_2) &  \dlin_\L(\vp_1)\dlin_\L(\vp_2) \,,\nonumber
\eea
are a correction to $\d$, where we notice that once more the first and second order prefactors match. 
The term in the second line of Eq.~(\ref{eq:shellGd}) reads
\bea
 4 \int_{\vp,\vp_1,\vp_2} \dirac(\vk-\vp_{12})  \s^2_{\vp,\vp_1}F_2(\vp_2,-\vp)  P_{\rm shell}(p) \dlin_\L(\vp_1)\dlin_\L(\vp_2) & \\
 = \int_{\vp,\vp_1,\vp_2} \dirac(\vk-\vp_{12}) \left[ -\frac{232}{105} + \frac{16}{105}\frac{(\vp_1\cdot \vp_2)^2}{p_1^2 p_2^2}  \right] & P_{\rm shell}(p) \dlin_\L(\vp_1)\dlin_\L(\vp_2) \,,\nonumber
\eea
and will contribute to $\d^2$ and $\G_2$. The term in the fourth line of Eq.~(\ref{eq:shellGd}) leads to
\bea
&&4 \int_{\vp,\vp_1,\vp_2} \dirac(\vk-\vp_{12}) \s^2_{\vp + \vp_{1},\vp_2}F_2(\vp,\vp_1)  P_{\rm shell}(p) \dlin_\L(\vp_1)\dlin_\L(\vp_2)  \\
&& =\quad \int_{\vp,\vp_1,\vp_2} \dirac(\vk-\vp_{12}) \left[ -\frac{52}{21} + \frac{20}{21}\frac{(\vp_1\cdot \vp_2)^2}{p_1^2 p_2^2}  \right]  P_{\rm shell}(p) \dlin_\L(\vp_1)\dlin_\L(\vp_2) \,,\nonumber
\eea
Finally
\bea
 \int_{\vp,\vp_1,\vp_2} \dirac(\vk-\vp_{12}) \s^2_{\vp + \vp_{1},\vp}F_2(\vp,\vp_1)  P_{\rm shell}(p) \dlin_\L(\vp_1)\dlin_\L(\vp_2) 
\eea
sources higher-derivatives due to $\s^2_{\vp + \vp_{1},\vp}$. Summing everything up, we find
\bea 
 \Shell^2_{\G_2\d}[\dlin_\L] (\vk) &=& \left[ - \frac{4}{3}\d^{(1+2)}(\vk) - \frac{376}{105}\d^{2,(2)}(\vk) +\frac{116}{105}\G_2^{(2)}(\vk) \right] \int \frac{p^2 dp}{2 \pi^2 }P_{\rm shell}(p) \\
 && \quad \quad \quad \quad \quad \quad \quad \quad \quad \quad \quad \quad \quad \quad \quad \quad \quad \quad + \mbox{h.d.} + \O\left[\left(\dlin_\L\right)^3\right]\,.\nonumber
\eea
    \item $O = \G_3$: Following the non-renormalization properties of Galileons \cite{Nicolis:2008in}, this term will only contribute to higher-derivatives as well \cite{Assassi2014}.

    \item $O = \Gamma_3$. Similar to the Galileon operator, all terms have a structure like $\s^2_{\vp + \dots,-\vp+\dots}$ or $F_3(\vp,-\vp, \vp_1)$:
{\small\bea
 \Shell^2_{\Gamma_3}[\dlin_\L] (\vk) &=&  4\dlin_\L(\vk) \int_{\vp}  \s^2_{\vk+\vp,-\vp} \left[F_2(\vk,\vp)-G_2(\vk,\vp)\right]  P_{\rm shell}(p) \\
 &+& 4 \int_{\vp,\vp_1,\vp_2} \dirac(\vk-\vp_{12}) \left[F_2(\vp,\vp_1)F_2(-\vp,\vp_2)-G_2(\vp,\vp_1)G_2(-\vp,\vp_2)\right] \s^2_{\vp_2-\vp,\vp+\vp_1} \vs
 && \hspace{8cm}\times P_{\rm shell}(p) \dlin_\L(\vp_1)\dlin_\L(\vp_{2}) \vs
  &+& 6 \int_{\vp,\vp_1,\vp_2} \dirac(\vk-\vp_{12}) \s^2_{\vp_1,\vp_2} \left[F_3(\vp,-\vp, \vp_1)-G_3(\vp,-\vp, \vp_1)\right] \vs
  && \hspace{8cm} \times P_{\rm shell}(p) \dlin_\L(\vp_1)\dlin_\L(\vp_{2}) \vs 
  &+&  6 \int_{\vp,\vp_1,\vp_2} \dirac(\vk-\vp_{12}) \s^2_{-\vp,\vk+\vp}  \left[F_3(\vp,\vp_1, \vp_2)-G_3(\vp,\vp_1, \vp_2)\right]  \vs 
  && \hspace{8cm} \times P_{\rm shell}(p) \dlin_\L(\vp_1)\dlin_\L(\vp_{2})\,, \nonumber
\eea}which only contributes to higher-derivative terms.

\end{itemize}
 
\section{Solution when including proxy for higher-order operators} \label{eq:proxisol}
We display here the solution for the ODE system Eqs.~(\ref{eq:drun_sigma})-(\ref{eq:Grun_sigma}) when taking the ansatz \refeq{bansatz} for the evolution of higher-order bias parameters: 
\bea
b_{\delta }(\s^2 ) &=& 
 b_{\d}^\ast +\frac{b^{*(\d)}_{n=3}\left(e^{-c^{(\d)} \s^2
   }-e^{-c^{(\d)} \text{$\s^2_\ast$}}\right)}{c^{(\d)}}  
   \\
   &+& \frac{e^{\text{$\s^2_\ast$} c_{\delta
   ^2,\delta ^2}} c_{\d,\delta ^2}
   \left[c^{(\d^2)} \left(b_{\delta^2}^\ast
   c^{(\d^2)}-b^{*(\d^2)}_{n=3+4} e^{-c^{(\d^2)}
   \text{$\s^2_\ast$}}\right) \left(e^{-\s^2 
   c_{\delta ^2,\delta ^2}}-e^{-\text{$\sigma_\ast$} c_{\delta ^2,\delta
   ^2}}\right)\right]}{c^{(\d^2)} c_{\delta
   ^2,\delta ^2} \left(c^{(\d^2)}-c_{\delta
   ^2,\delta ^2}\right)} 
   \vs
   &+& \frac{c_{\d,\delta ^2}
   \left[c_{\delta ^2,\delta ^2}
   \left( b^{*(\d^2)}_{n=3+4} e^{-c^{(\d^2)}
   \s^2 }-b^{*(\d^2)}_{n=3+4} e^{-c^{(\d^2)}
   \text{$\s^2_\ast$}}-b_{\delta^2}^\ast
   c^{(\d^2)}+b_{\delta^2}^\ast c^{(\d^2)} 
   e^{(\text{$\s^2_\ast$}-\text{$\s^2$}) c_{\delta
   ^2,\delta ^2}} \right)\right]}{c^{(\d^2)} c_{\delta
   ^2,\delta ^2} \left(c^{(\d^2)}-c_{\delta
   ^2,\delta ^2}\right)}  \,, 
    \vs
b_{\delta ^2}(\s^2 ) &=&  \frac{e^{-(\s^2 -\text{$\s^2_\ast$}) c_{\delta^2,\delta ^2}} \left(b_{\delta^2}^\ast c_{\delta^2,\delta^2}+b^{*(\d^2)}_{n=3+4} e^{-c^{(\d^2)}
   \text{$\s^2_\ast$}}-b_{\delta^2}^\ast
   c^{(\d^2)}\right)-b^{*(\d^2)}_{n=3+4} e^{-c^{(\d^2)}
   \s^2 }}{c_{\delta ^2,\delta
   ^2}-c^{(\d^2)}} \,, \label{eq:d2sol_2}
   \\
b_{\G_2}(\s^2 ) &=&  b_{\G_2}^\ast +\frac{b^{*(\G_2)}_{n=3+4}\left(e^{-c^{(\G_2)} \s^2
   }-e^{-c^{(\G_2)} \text{$\s^2_\ast$}}\right)}{c^{(\G_2)}}  \label{eq:Gsol_2}
   \\
   &+& \frac{e^{\text{$\s^2_\ast$} c_{\delta
   ^2,\delta ^2}} c_{\G_2,\delta ^2}
   \left[c^{(\d^2)} \left(b_{\delta^2}^\ast
   c^{(\d^2)}-b^{*(\d^2)}_{n=3+4} e^{-c^{(\d^2)}
   \text{$\s^2_\ast$}}\right) \left(e^{-\s^2 
   c_{\delta ^2,\delta ^2}}-e^{-\text{$\sigma_\ast$} c_{\delta ^2,\delta
   ^2}}\right)\right]}{c^{(\d^2)} c_{\delta
   ^2,\delta ^2} \left(c^{(\d^2)}-c_{\delta
   ^2,\delta ^2}\right)} 
   \vs
   &+& \frac{c_{\G_2,\delta ^2}
   \left[c_{\delta ^2,\delta ^2}
   \left( b^{*(\d^2)}_{n=3+4} e^{-c^{(\d^2)}
   \s^2 }-b^{*(\d^2)}_{n=3+4} e^{-c^{(\d^2)}
   \text{$\s^2_\ast$}}-b_{\delta^2}^\ast
   c^{(\d^2)}+b_{\delta^2}^\ast c^{(\d^2)} 
   e^{(\text{$\s^2_\ast$}-\text{$\s^2$}) c_{\delta
   ^2,\delta ^2}} \right)\right]}{c^{(\d^2)} c_{\delta
   ^2,\delta ^2} \left(c^{(\d^2)}-c_{\delta
   ^2,\delta ^2}\right)} \,. \nonumber
\eea

\bibliographystyle{JHEP}
\bibliography{main}

\end{document}